\documentclass[]{mn2e}  
\input epsf                                          
\usepackage[draft]{graphics}                  
\usepackage[]{color}
\title[Rotation effects on pulsar intensity profiles]{Understanding the 
effects of geometry and rotation on pulsar intensity profiles. }

\author[R.M.C. Thomas, Y. Gupta and R.T. Gangadhara]
{R.M.C. Thomas $^{1}$\thanks{E-mail:
mathew@ncra.tifr.res.in}  Y. Gupta $^{1}$\thanks{E-mail: ygupta@ncra.tifr.res.in} and
 R.T. Gangadhara$^{2}$\thanks{E-mail: ganga@iiap.res.in }
\\
$^{1}$National Center for Astrophysics, Pune 411007, India\\
$^{2}$Indian Institute of Astrophysics, Bangalore 560034, India}
\begin{document}

\date{ Date of acceptance to be inserted}

\pagerange{\pageref{firstpage}--\pageref{lastpage}} \pubyear{2002}

\maketitle 
 
\label{firstpage}

\begin{abstract}

We have developed a method to compute the possible distribution of
radio emission regions in a typical pulsar magnetosphere, taking into
account the viewing geometry and rotational effects of the neutron
star.  Our method can estimate the emission altitude and the radius of
curvature of particle trajectory as a function of rotation phase for a
given inclination angle, impact angle, spin-period, Lorentz factor,
field line constant and the observation frequency. Further, using
curvature radiation as the basic emission mechanism, we simulate the
radio intensity profiles that would be observed from a given
distribution of emission regions, for different values of radio
frequency and Lorentz factor.  We show clearly that rotation effects
can introduce significant asymmetries into the observed radio
profiles.  We investigate the dependency of profile features on
various pulsar parameters.  We find that the radiation from a given
ring of field lines can be seen over a large range of pulse
longitudes, originating at different altitudes, with varying spectral
intensity.  Preferred heights of emission along discrete sets of field
lines are required to reproduce realistic pulsar profiles, and we
illustrate this for a known pulsar.  Finally, we show how our model
provides feasible explanations for the origin of core emission, and
also for one-sided cones which have been observed in some pulsars.
\end{abstract}
  
\section{Introduction}

Of the various aspects relevant for solving the unresolved problem of
radio emission from pulsars, two are probably the most significant :
the actual mechanism of the emission itself, which is still not fully
understood (e.g. Zhang 2006, Melrose 2006); and the effects of viewing
geometry and pulsar rotation, which can significantly alter the
properties of the pulsar profiles that the observer finally samples.
The latter aspect has received significant attention in the recent
years, but much still remains to be investigated.  Blaskeiwicz
et~al. (1991, hereafter BCW91) were the first to work out the basic
effects of rotation, and showed that the observed asymmetry between
the leading and trailing parts of pulsar radio profiles can be due to
rotation effects.  Further improvements were carried out by Hibschman
\& Arons (2001), who analyzed the first order effects of rotation on
the polarization angle sweep.  Later, Peyman \& Gangadhara (2002),
adapting the method of BCW91, refined the formulation and showed that
the asymmetries due to rotation can be ascribed to the differences in
the radius of curvature of the particle trajectories on the leading and
trailing sides of the magnetic axis.  By analysing the pulse profiles
of some selected pulsars, which clearly show the core-cone structure in
the emission beam, Gangadhara \& Gupta (2001, hereafter GG01) and Gupta
\& Gangadhara (2003, hereafter GG03) showed that the asymmetry in
the locations of the conal components around the central core
component can be interpreted in terms of aberration and retardation
(A/R) effects (combined effects of rotation and geometry), leading to
useful estimates of emission heights of the conal components.  Further
refinements of these concepts have been carried out by Dyks et~al. (2004),
and Gangadhara (2004 \& 2005, hereafter G04 \& G05), Dyks (2008) and
Dyks et~al. (2009).
 
All of the above said works have established that rotation effects 
are of significant importance in understanding the observed emission 
profiles of radio pulsars.  What has been found wanting is a detailed, 
quantitative treatment that couples the  rotation effects in the pulsar 
magnetosphere to the possible emission physics and to the emission and 
viewing geometries, to produce observable radio profiles.  Some 
impediments to this have been recently overcome by Thomas and Gangadhara
(2007, hereafter TG07), who have considered in detail the dynamics of 
relativistic charged particles in the radio emission region, and obtained
analytical expressions for the particle trajectory and it's radius of 
curvature. 
      
In this paper, we describe a scheme for simulation of pulsar profiles 
that encompasses a detailed treatment of all the effects mentioned 
above.  We start with describing the background and motivation for 
the work (\S\ref{sec:significance}), then go on to the profile 
simulation method (\S\ref{sec:prof}).  We describe the main results 
from our study and their dependence on pulsar parameters in
\S\ref{sec:results-discussion}, and discuss how realistic pulsar
profiles may be obtained from our model.  We also address the issues
of core emission, one-sided or partial cones, and extension of our
method to other models of emission physics.  Our final conclusions are
summarized in \S\ref{sec:summary}.
   
\section{Significance of geometry and rotation effects} \label{sec:significance}
The charged particles produced in the pulsar magnetosphere are
initially accelerated in a region very close to the polar cap, 
due to the electric fields generated by the rotating magnetic field 
(Sturrock 1971; Ruderman \& Sutherland 1975; Harding \& Muslimov 1998). 
After crossing the initial acceleration region they enter the 
radio-emission domain where the parallel component of the
electric field is screened by the pair plasma, and henceforth they
move `force-free'.  In the rotating frame, the charged particles are
constrained to move along the field lines of the super-strong magnetic
field, the geometry of which is believed to be predominantly dipolar
in the radio emission region (e.g., Xilouris et~al. 1996; Kijack \& Gil
1997) -- the multi-polar components of the pulsar  magnetic field are
expected to be limited to much lower altitudes close to the stellar
surface.  The accelerated charges are believed to produce coherent
radio emission at specific altitudes in the magnetosphere, by a
mechanism that is as yet not fully understood (e.g. Ginzburg et~al. 1969;
Ginzburg \& Zheleznyakov 1975; Melrose 1992a, 1992b; Melrose 2006).

Though the trajectory of the particles in the co-rotating frame is
identical to that of the field line they are associated with, the
trajectory will be significantly different in the observer's frame,
due to the effect of rotation (TG07).  The velocity vector of the
particles will be offset from the tangent vector of the field line on
which they are constrained to move.  The value of the azimuthal angle
of this offset is termed as the aberration phase shift $\delta\phi_{\rm
  aber} $ which depends on the emission altitude and also on
inclination angle $\alpha$ and impact angle $\beta$ (G05).  One
consequence of this rotation effect is that the radius of curvature
$\rho$ of the particle trajectory becomes significantly different from
that of the field line, as shown clearly by TG07.  This can be
intuitively understood as follows : on the leading side, the induced
curvature due to rotation has the same sense as the curvature of
the field lines and hence the net curvature of the particle trajectory
in the observer’s frame gets enhanced, resulting in a reduced value of
$\rho$, as compared to that for the corresponding field line.  On the
trailing side, the curvature of the field lines and the induced
curvature due to rotation are in opposite directions, and hence they
counter-act to result in a reduced effective curvature (or a larger
value of $\rho$) for the particle trajectory.  It was shown in TG07
that the disparity between the values of radii of curvature of field
lines and that of the corresponding particle trajectory could be
substantial.  For example, for a pulsar with $\alpha=90^{\circ}$
  and spin period $P=1$~sec, at an emission altitude of 0.04 of the
  light cylinder radius ($r_{\rm L}$), along a field line with field
  line constant $r_{\rm e}=50,$ the ratio of the $\rho$ estimated with
  and without rotation effects is more than $ 4$.  At an altitude 0.08
  $r_{\rm L},$ the ratio is more than $ 5$ (see Fig.~6 in TG07).
  These ratios steeply increase for inner field lines.  and indicate
  that the rotation induces substantial and unavoidable differences on
  curvature radii and intensity of emission between leading and
  trailing sides.  Further, it was shown in TG07 that the maximum
  value of the $\rho$ for the particle trajectory, corresponding to
  field lines either very close to magnetic axis or the ones falling in
  the meridional plane, including the effects of rotation can be
  $\rho_{\rm max}=r_{\rm L}/(2\sin\alpha).$ While the maximum of $\rho$
  for these field lines in the non-rotating case can attain any value
  up to infinity.  Thus, a proper estimation of $\rho$ for the
particle trajectory needs to take into account the effects of
rotation.  Unfortunately, this consideration has been missing in
several works where  $\rho$ has been presumed to be identical to
that of the dipolar field lines (e.g. Cheng \& Zhang 1996; Lyutikov
et~al 1999; Gil et~al. 2004).

The effects of rotation and pulsar geometry can combine to produce
interesting results in observed pulsar profiles, the details of
which can depend somewhat on the specific models for the emission
process, including that for the distribution of regions producing
accelerated charged particles on the polar cap.  For our immediate
purposes, we use the basic model of nested cones of emission, along
with a possible central core, as postulated and demonstrated by
several authors (e.g., Rankin 1983, 1993a, 1993b; Mitra \& Deshpande 1999;
GG01; GG03).  In such models, the source of emission could
be in the from concentric rings of sparks produced in the polar
vacuum gap, and circulating around the magnetic axis (e.g., Gil \&
Krawczyk 1997, Deshpande \& Rankin 1999, Gil \& Sendyk 2003).  Each
ring or cone is represented by a narrow annulus of field lines
characterised by a definite value of the field line constant
$r_{\rm e}$ appearing in the field line equation $r = r_{\rm e}
\sin^2\theta$, where $r$ is the radial distance to an arbitrary
point on the field line and $\theta$ is the magnetic co-latitude. 
The set of field lines can also be identified by $S_{\rm L}$, the 
distance from the magnetic axis, of the point where the field line 
pierces the neutron star surface, normalized to that for the last 
open field line (GG01).

In the above model, if we assume that the charged particles along
a given conal set of field lines emit radiation at a given height in
the co-rotating magnetosphere, then the simplest effect of rotation
and pulsar geometry is to shift this radiation by $\delta\phi_{\rm
 aber}$ in pulse longitude on the leading and trailing sides of the
profile (with respect to the magnetic meridian).  This is because
rotation causes the emission beam to be offset from the local field
line tangent in the observer's frame, hence causing the
corresponding emission component to be advanced in azimuthal phase
by the same amount.  This is the well understood aberration effect
which, in combination with the retardation effect, leads to phase
asymmetry between the leading and trailing side components associated
with a given cone of emission, and has been explored in detail by
several authors (e.g., GG01; GG03; Dyks \& Harding 2004; Dyks
et~al. 2004; G05).  
      
Furthermore, if we assume that the emission
mechanism is such that the curvature of the particle trajectory
plays an important role in the generation of the radio waves, as
would be the case for models related to the curvature radiation
mechanism, then rotation effects can produce significant changes in
the strength of the intensity profiles on the leading and trailing
sides.  This has been indicated by several authors (e.g., BCW91;
Peyman \& Gangadhara 2002; TG07; Dyks 2009).  For example, TG07 have
shown quantitatively that the ratio of the total intensity estimated
for the same field line with and without rotation effects at an
emission altitude of 0.04 $r_{\rm L}$ is more than a factor of $13,$
and at 0.08 $r_{\rm L}$ it is more than a factor of $30$ (see Fig.~8
in TG07).  In extreme cases, this effect could lead to almost
one-sided intensity profiles.  That such effects are seen in
profiles of known pulsars (e.g., Lyne \& Manchester 1988, hereafter
LM88) is a strong indicator of the importance of rotation effects.

All this motivates the importance for a method that can estimate the 
properties of the received emission, after including the effects 
described above.  The necessity of such a 3D method is pointed out 
by Wang et~al. (2006).  A good way to proceed is to simulate the 
emission properties for a given choice of pulsar parameters and predict 
the observed profiles that would be seen for a range of the parameter 
space.  These can then be compared with realistic pulsar profiles to 
gain a better understanding of the physical processes involved.  We
have developed such a scheme, which is described in detail in the
following sections.

\section{Profile simulation studies}\label{sec:prof}

\subsection{Basic concepts}

The ultra-relativistic particles that are constrained to move along
the co-rotating magnetic field lines suffer acceleration and hence
emit beamed radiation within a narrow angular width of $2/\gamma$,
centered on the direction of the instantaneous velocity vector, $\bf
v$.  This emission is aligned with the local tangent vector $\bf b$ to
the field line in the co-rotating reference frame. However, as
mentioned earlier, in the observer's reference frame, the velocity
vector is offset from the field line tangent vector (GG01, G05).
Furthermore, the radius of curvature of the particle trajectory will
be different from that of the associated field lines (TG07).  To begin
with, for a given pulsar geometry we choose a ring of field lines
specified by a single value of field line constant $r_{\rm e}$ (or by
the equivalent value of $S_{\rm L}$) and look for all possible emission 
spots along the field line that can contribute in the observer's 
line-of-sight direction.  For the observer to receive significant 
radiation, we impose the condition that the unit velocity vector,
$\hat\textbf{v}$, should align with the unit vector along the 
line-of-sight, $\hat\textbf{n}$, such that
\begin{equation}\label{eq_los_vel}      
{\hat \textbf n}\cdot {\hat \textbf v}=1  ~~.
\end{equation} 
For any given pulse phase of observation,
we find the possible emission spots on the specified ring of field
lines that meet this criterion.  For each emission spot, we compute
the emission altitude, $r$, the emission angles $\theta$ and $\phi$ 
(defined in sec. ~\S\ref{sec:details}), and the radius of curvature of 
the particle trajectory, $\rho$. 

We then couple the basic curvature radiation model to the above picture,
using the value of $\rho$ in order to obtain estimates of the specific 
intensity that would be seen by the observer at a given pulse longitude.    
Of the several emission mechanisms proposed for pulsar emission, the 
curvature emission model is perhaps the most natural and favoured one 
(Gil et~al. 2004).  However, as we argue later, our method is flexible 
enough to incorporate other viable variants of pulsar emission models.  
Using the curvature radiation formulation, we compute the intensity that
would be seen at $\it any$ observing frequency, $\omega$, and not
just for the characteristic frequency, $\omega_{\rm c}.$  We sweep the
line of sight through discrete rotation phases and, for any given
field line, search for the points that contribute at a given
$\omega$, and calculate the emission parameters like emission height,
radius of curvature and, finally, the intensity of the radio emission
received.  This method allows us to generate a super-set of all
possible emission profiles that would be observed at a given
frequency.
  
\subsection{Details of the method}\label{sec:details}

\begin{figure}  
\begin{center}
\epsfxsize= 7 cm
\epsfysize=0cm
\rotatebox{0}{\epsfbox{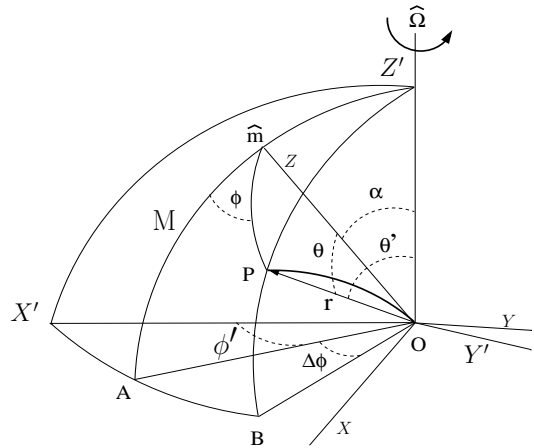}}
\caption[short_title]{\small A schematic picture that depicts the 
geometry associated with the emission region. The $XYZ$ frame is
co-rotating with the magnetic axis ${\hat m}$ around spin axis
${\hat\Omega},$ and the $X'Y'Z'$ frame is the non-rotating
frame of the observer.  
\label{fig_GEO}}
\end{center} 
\end{figure}

Any emission point in the pulsar magnetosphere can be located by the
coordinates $r,$ $\theta$ and $\phi$ in a coordinate system where the
$Z$--axis is parallel to the magnetic axis $\hat \textbf{m}$, and the
$XZ$--plane is the plane containing the magnetic and rotation axes
(see Fig.~\ref{fig_GEO}).  This coordinate system can be called as the
magnetic coordinates, where $\theta$ and $ \,\phi$ are the magnetic
co-latitude and the magnetic azimuth, respectively.  This
coordinate system co-rotates with the pulsar.  Another coordinate
system, identified by $X'Y'Z'$ (the observer's frame), can be defined 
such that the line-of-sight vector is parallel to the $X'Z'$--plane 
containing the magnetic axis and the rotation axis at rotation phase 
$\phi'=0,$ and designated as the meridional plane $M.$  The $Z'$--axis 
is parallel to the rotation axis $\hat \Omega$.  The {\bf $X'Z'$--plane} 
makes an azimuthal angle $\phi'$ with $M$ during rotation.

In the $X'Y'Z'$ coordinate system, the location of the emission spot 
can be specified by the values of $r,$ $\theta'$ and $\phi'_{\rm p}$ 
(see \S\ref{app:vel}).  The expressions for $\theta$ and $ \phi$
are given in G04, and the expression for $ \theta'$ as a function of the
pulsar rotation phase $\phi'$ is given in G05.  A point of emission on
a dipolar field line can be expressed as $r=r_{\rm e} \sin^2\theta$.
When rotation effects are not included, it turns out to be a trivial
exercise to trace emission spots based on the expressions for $r(t),$ 
$\theta$ and $\phi$.  For a given $\alpha$, $\beta$ and $S_{\rm L}$ 
combination, the value of $r(\phi') = r_{\rm e}\sin^2[\theta(\phi')]$, 
the radius of curvature $\rho$ can be readily found (Eq. 4 in G04).  
However, the estimation of $r$ and $\rho$ corresponding to the 
emission spot becomes difficult when the effects of rotation are 
included.  
   
As mentioned before, when rotation is invoked, the observer will
receive peak radiation when the particle velocity vector ${\hat \textbf v}$
(rather than the tangent to the field line) becomes parallel to the
line-of-sight $\hat n$, where ${\hat \textbf {n}}=[\sin\zeta,0,\cos\zeta]$, 
and $\zeta=\alpha+\beta$.  The total velocity of the charged particle
will be the vector sum of the component parallel to the magnetic
field, and the component in the azimuthal direction due to the
co-rotation of the field (TG07).  The component in the azimuthal
direction will make the total velocity offset from the direction of
the local field tangent.  The analytical solution for the radial
position of the trajectory of the charged particle, including the
rotation effects, is derived in TG07.  We employ the zeroth order
solution that gives :
\begin{equation}\label{eq_solution_0_order}
r(t)  =  \frac{c}{\Omega_{\rm m0} }\rm{cn}(\lambda-\Omega_{\rm m0}\, t )
\end{equation}       
where $\Omega_{\rm m0} = \Omega \sin\alpha, $ and $\rm cn(z)$ is the
Jacobi Cosine function (Abramowitz \& Stegun 1972), $\lambda$ is a
constant, $t$ is the affine time and $c$ is the speed of light.

For a given $\alpha,$ $\beta$ and a rotation phase $\phi'$, the
radial distance for an emission spot can, in principle, be found by
solving Eq.~(\ref{eq_los_vel}).  The rotation phase $\phi'$ is defined
as the projected angle on the azimuthal plane between ${\hat \textbf {n}}$ 
and $\hat \textbf {m} $ (see Eq.~(\ref{eq_rot_phase}) in \S A).  The 
above-said analytical expressions for the coordinates of the 
accelerated charged particle are to be as invoked as functions 
of affine time.  The aberration phase shift $\delta\phi_{\rm aber}$ 
needs to be known in advance for solving Eq.~(\ref{eq_los_vel}) and 
finding the emission altitude, since the angles $\theta$ and $\phi$ 
are aberrated due to the rotation (see G05 for details, including the 
analytical expression for $\delta\phi_{\rm aber}$).  Nonetheless, the 
emission altitude is apriori required for estimating the 
$\delta\phi_{\rm aber}$, thus making the problem non-linear. An 
analytical solution for Eq.~(\ref{eq_los_vel}) is nearly impossible 
owing to the bulky trigonometric terms.  Approximations like 
$\delta\phi_{\rm aber} \approx r/r_{\rm L}$, will severely limit
the expected precision of the solution.  A straight forward numerical
solution for Eq.~(\ref{eq_los_vel}) also encounters more or less 
similar difficulties owing to the aforesaid reasons.  Hence we have 
developed special algorithms which are applicable for solving 
Eq.~(\ref{eq_los_vel}) under such conditions.  We have devised an 
`exact' method and also an `approximate' method to compute the 
location of possible emission spots, which are suited for specific 
parameter regimes.  These are described below.
       
\subsubsection{The `exact' method}\label{sec:exact}
For a ring of field lines specified by a field line constant
$r_{\rm e}$ and for a given rotation phase $\phi'$, we consider a
point $P_0$ on a field line such that the unit-vector of the local
field line tangent $\hat \textbf{b}_0$ is parallel to $\hat \textbf{n}$ 
in the {\it non-rotating case}.  In the absence of rotation, the 
emission beam from the accelerated particles moving along the
field line should be aligned with $\hat \textbf{b}_0$. But when the
effects of rotation are invoked, the emission beam at $P_0$ gets 
mis-aligned with $\hat \textbf{b}_0$ and goes out of the
line-of-sight. Hence the radiation from $P_0$ will not be received
by the observer.  However, another emission spot $P_1$ on the same
ring of field lines can have the $\hat\textbf{v}$ parallel to
$\hat\textbf{n}$ and contribute emission in the direction of the
observer.  Hence the observer will receive radiation from $P_1$
provided Eq.~(\ref{eq_los_vel}) is satisfied for $P_1.$  Let the
aberration phase shift at $P_1$ be $\delta\phi_{\rm aber}$; then
the updated values $\theta=\theta(\phi'+\delta\phi_{\rm aber})$ and
$\phi=\phi(\phi'+\delta\phi_{\rm aber})$ at the point $P_1$ so that
the emission is aligned with the line of sight to the observer (G05).

Hence the basic idea that is invoked in the computation of the exact
method can be briefed as: for a given $\phi',$ begin with the value of
the emission height, which is estimated in the non-rotating case, and
find the trial value of $\delta\phi_{\rm aber}$. Then solve
Eq.~(\ref{eq_los_vel}) numerically to find an improved value of
emission height.  Continue the iteration till the solution for
Eq.~(\ref{eq_los_vel}) satisfactorily converges.  The main steps of
algorithm are briefly described below :
   
\begin{enumerate}
\item{Choose  a specific combination of 
  $\alpha,$ $\beta,$ $\gamma,$ $r_{\rm e}$ and $\omega, $ and  a fixed
  rotation phase $\phi'.$ }
\item{Make the first estimate for the aberration angle
  $\delta\phi_{\rm aber}$ using the trial input values of $\theta(\phi'),$
  $\phi(\phi')$  and $r,$ which follows from $r=
  r_{\rm e} \sin^2\theta(\phi')$.}
\item{Using the $\delta\phi_{\rm aber}$ estimated above, the angles
  $\phi$ and $\theta$ are re-calculated with the rotation phase
  incremented by $\delta\phi_{\rm aber}.$  Henceforth update: $\phi
  (\phi') \rightarrow \phi (\phi'+ \delta\phi_{\rm aber}) $ and
  $\theta(\phi') \rightarrow \theta (\phi'+ \delta\phi_{\rm aber})
  $. }\label{step_dphi} 
\item{ Estimate $\bf v$ and hence the unit vector ${\hat \textbf{v}}={\bf
  v}/|{\bf v}|$ with the angles $\phi$ and $\theta,$ found  in
  step \ref{step_dphi}. } 
\item{ Estimate the affine time `t' that satisfies the matching
  condition ${\hat \textbf{n}}\cdot{\hat {\bf v}}=1$. Hence
  find the improved value of $r(t).$}
\item{Recalculate $\delta\phi_{\rm aber}$ with  $r=r(t)$ and
  repeat the calculation from step~\ref{step_dphi} till convergence is
  achieved for `t'.}
\item{Using the improved value of $r(t)$ find $\bf v$, $\bf a,$ and
  $\rho.$ }
\item{ Using the $\rho$ estimate the  spectral intensity. }
\item{ Find the angle $\eta_{\rm mis}=\cos^{-1}({\hat
   \textbf  n}\cdot{\hat \textbf {v}}).$ }   
\end{enumerate}  
      
We choose to call this method as `exact method' since the computation
employs the exact expressions for the relevant quantities.  The
explicit expressions for magnetic co-latitude $\theta$ and magnetic
azimuth $\phi$ are given by Eq.~(\ref{eq_mag_azimuth}) and 
Eq.~(\ref{eq_theta_dash}), respectively, in \S A.  The expressions 
for velocity $\textbf v$ and acceleration $ \textbf a$ are given 
by Eq.~(\ref{eq_par_tot}) and Eq.~(\ref{eq_par_acc}) respectively, 
and the expression for radius of curvature $\rho$ is given in 
Eq.~(\ref{eq_rho}).  Sample results from this method are shown in 
the figures in \S\ref{app:figs}.  The angle $\eta_{\rm mis}$ gives 
the residual difference between the line-of-sight and the estimated 
${\textbf v}$ at the end of the iterations.  It's ideal value is zero
and hence, the final residual value obtained is a measure of the 
precision of the solution : a smaller value of $\eta_{\rm mis}$ 
indicates a more precise determination of the emission spot.

\subsubsection{Alternative or `approximate' method}\label{sec:approx}
 
We have devised an alternative or approximate method, for the
estimation of $r$, $\rho$ and the related quantities, in cases 
where we encounter `extreme' values of parameters.  Such regimes
are often combination of large values of $\alpha$, very low values 
of $\beta,$ and field lines close to magnetic axis ($S_L < 0.5)$.  
The exact method encounters difficulties for such regimes in that 
the numerical solutions of Eq.~(\ref{eq_los_vel}) for affine time 
`t' often do not give satisfactory convergence. So we resort to an 
approximate method that is suitable for this regime.  By this method, 
we expect to determine the emission height and the radius of
curvature with comparable precision to the exact one, for leading and
the trailing parts of the pulsar profile.  The estimates of this
method have been optimised by comparison with the estimates of the
aforesaid exact method in a common parameter regime where both the
methods give reliable results. The scheme of the approximate method
is: for a given rotation phase $\phi',$ first calculate $r,$ $\theta$
and $\phi$ in the {\it non-rotating case,} and then use it to find
approximate values of $r,$ $\theta$ and $\phi$ in the {\it rotating
  case}.  The details are provided in \S\ref{app:approx}.  
      
\subsection{Computing the intensity profiles}\label{sec:comp-intensity} 
 
As explained earlier, we sweep the line of sight through discrete
rotation phases, and using the afore-said steps given in 
\S\ref{sec:exact} and \S\ref{sec:approx}, we find the parameters 
like emission altitude $r$, and the radius of curvature of the particle 
trajectory, $\rho$.  Then we estimate the spectral intensity for a 
given frequency, $\omega$, for particles for a given Lorentz factor, 
$\gamma$, by using the standard curvature radiation formula 
(e.g. Jackson 1972) :
\begin{equation}\label{eq_spec_Int} 
d\,I/d\omega ~=~ \sqrt{3} \frac{e^2}{c} \gamma \,\, u 
\int^{\infty}_{u} K_{5/3}(u') du'~, 
\end{equation}
where $u={\omega}/{\omega_{\rm c}}$ and the characteristic frequency
$\omega_{\rm c} = 1.5\,\gamma^3\,c/\rho$.  According to
Eq.~(\ref{eq_spec_Int}), the spectral-intensity curve should peak 
at $u=0.286$, which corresponds to $\rho/\rho_{\rm p}=1$, where the
parameter $\rho_{\rm p} $ can be defined as 
\begin{equation} \label{eq_rho_p}
\rho_{\rm p} = 0.286 \times 1.5 \gamma^3\frac{c}{\omega}~.
\end{equation}
Invoking this parameter helps in easy identification of the peak 
points in the spectral-intensity plots.

An important feature of this method is that it computes the
contribution to the observed intensity for any frequency $\omega,$
different from $\omega_{\rm c}.$  Often in literature, significant
contribution of intensity to the observer is presumed to be
concentrated near the characteristic frequency $\omega_{\rm c}= 1.5
\gamma ^3 c/\rho$, thus providing an in-built frequency selection
criteria (e.g., Melrose 2006).  However, the spectral intensity 
curve for curvature radiation has non-negligible amount of power 
emitted at a significant range of frequencies different from 
$\omega_{\rm c}.$  Our formulation thus 
allows for a more complete treatment of the amount of emitted 
intensity and its reception by the observer.
     
Since our present analysis necessitates only the computation of relative
intensities of the simulated profiles, invoking a single particle
emission model for the curvature emission do not alter the results in
a significant manner. The high luminosity of the pulsars demands
imposing coherence on the emission, perhaps in the form of bunched emitting sources,
and the process behind the formation of such bunches is still being
investigated. In a simple manner, coherent emission from a bunch with
charge Q can be alternatively expressed as the emission from a single
particle with the same charge Q.  So, the  relative intensities are
not affected by this simplification  and hence considering single
particle emission do not tamper with the physics behind the emission.
Further discussions regarding this factor will be followed in  later
sections.

\subsection{Typical outputs}  
We have computed the parameters of emission in the magnetosphere by
implementing the method described above.  The free parameters are
the following : $\alpha$, $\beta$ (pulsar geometry), $\Omega$ (pulsar 
rotation frequency), $\gamma$ (Lorentz factor of the particles),
$S_{\rm L}$ (field line location) and $\omega$ (radio frequency of 
observations).  For the sake of brevity of presentation, we give 
results for a single fixed value of $\Omega = 2\pi$ (i.e. a spin period 
of 1 sec), and for a relatively narrow range of rotation phases of 
about $-10^{\circ}$ to $+10^{\circ}$ around zero (fiducial) phase.  
As is discussed later, frequency turns out to be a relatively weak 
parameter in comparison to other strong ones that influence profile 
evolution.  Hence, for the simplicity of analysis, we have restricted 
the frequency to a single value of 610 MHz.  For a chosen pulsar geometry, 
the emission locations are estimated for each discrete rotation phase 
and for a set of discrete choices of $S_{\rm L}.$  Further, the specific 
intensity values are estimated for a set of discrete values of $\gamma$, 
for a fixed value of $\omega.$

The typical outputs are shown in the figures in \S\ref{app:figs}.
Fig.~\ref{fig_A30-B1} shows the basic outputs from a typical
simulation run for estimating the location of emission regions, for
a fixed pulsar geometry, for a set of $S_{\rm L}$ values (0.1, 0.3, 
0.5 and 0.7). The following quantities are shown, as function of 
rotation phase, in separate panels for each $S_{\rm L}$ value : the 
estimated emission altitude, $r$; the computed radius of curvature 
of the particle trajectory, $\rho$; the $\phi$ and $\theta$ for the 
emission spots in the magnetosphere; and the mis-alignment angle, 
$\eta_{mis}$, which is a measure of the accuracy of the results. 
Fig.~\ref{fig_A30-B1-SPEC} shows the computed intensity profiles for 
each choice of $S_{\rm L}$ values (in separate panels) for a set of 
$\gamma$ values (200, 300, 400, 600, 1000, 1500), for a fixed 
observation frequency of 610 MHz.  
These two figures illustrate the basic results. The effect of varying 
pulsar geometry can now be explored to understand the variety of 
intensity profiles that are possible.  Most important outputs are 
$r/r_{\rm L}$, $\rho/r_{\rm L}$ and specific intensity plots, and 
the successive figures show only these quantities.

Figures~\ref{fig_A30-B2-ALL} to \ref{PARTIAL} then investigate the
effects as a function of pulsar geometry (i.e. different combinations
of $\alpha$ and $\beta$ values), with the range of $S_{\rm L}$ values and
$\gamma$ values confined to those illustrated in Figs.~\ref{fig_A30-B1}
and \ref{fig_A30-B1-SPEC}.

\section{Results and discussion}\label{sec:results-discussion}
\subsection {General results}

We first discuss the general results and trends that are deduced from 
our simulation studies, as illustrated in the results displayed in the 
plots given in \S\ref{app:figs}.  

\subsubsection{Emission heights} 

The heights for the allowed emission spots have a minimum value near
the magnetic meridian ($\phi'=0$), with smoothly increasing values on
the leading and trailing sides.  However, the variation of height with
rotation phase is asymmetric such that the  increment with rotation phase 
is always faster for the trailing side than for the leading side.
Whereas this increase with rotation phase is purely geometric, the
asymmetry in this is due to the modification of particle trajectories
produced by rotation.  On the leading side, the emission beam bends in 
the direction favourable to rotation and hence advances in azimuthal 
phase by $\delta\phi'_{\rm aber}$ from that of the corresponding field 
line tangent. Hence at a fixed $\phi',$ an emission spot located at a 
lower emission height than in the non-rotating case will satisfy 
Eq.\ref{eq_los_vel}.  In contrast, on the trailing side, the bending 
of the emission beam causes a lag in azimuthal phase by 
$\delta\phi'_{\rm aber}$ from the corresponding field line tangent. 
This lag can be compensated by an emission spot located at a different 
altitude than the non-rotating case.  Hence an emission spot located 
at a  different emission height will satisfy Eq.\ref{eq_los_vel} and 
contribute radiation to the observer. 
  
Both the value of the minimum height (at $\phi'=0$), as well as the
asymmetry, are larger for the inner field lines as compared to the
outer field lines.  This supports the intuitive expectation that outer
field lines would be visible out to much larger pulse longitude ranges
than inner field lines.  Further, it is seen that the minimum height
and asymmetry of the variation increase with geometry, being more for
larger values of $\alpha$ and $\beta$.  Also, pulsars with larger
values of $\alpha$ will have a larger range of variation of allowed
emission heights, for the same field line. It is reasonable to surmise
that a similar value of emission height can be seen recursively for several
combinations of larger and smaller values of $\alpha$ and $S_{\rm L},$
for a certain $\phi'$.  Thus we find that the values of $\alpha$ and
$S_{\rm L}$ dominantly decide the range of emission heights. Even
otherwise, the $\alpha$ dependence of the emission height can be
directly understood from Eq.(\ref{eq_solution_0_order}), since the
expression $r(t)$ is explicitly dependent on the value of $\alpha.$

\subsubsection{Radius of curvature} 

The radius of curvature inferred for the possible emission spots also
varies signficantly with rotation phase, and it is significantly
asymmetric between leading and trailing sides.  If rotation effects
were not considered, this radius of curvature would be same as that of
the corresponding field line and  be symmetric around zero pulse
phase.  Given that the observed radius of curvature of the particle
trajectory is a combination of the curvature of the field line and
curvature introduced due to rotation, the observed asymmetry is a
rotation effect and can be understood as a combination of two effects:
first, as the allowed heights are different on the leading and
trailing sides (for the same phase $\pm \phi'$ on either side of the
zero phase), the radius of curvature of the field line itself would be
different; second, on the leading side, the curvature of the field
line gets combined with that induced by rotation (TG07), resulting in a
lower value of the $\rho$ for the particle trajectory.  Whereas on the
trailing side, the two curvatures are opposed and hence the net
curvature is reduced, leading to larger values of $\rho.$ In fact, for
some field lines, rotation induced curvature can cancel the curvature
due to the field lines, at some points on the trailing side, resulting
in sharply peaked curves for $\rho$ for certain values of $S_{\rm L},$
as seen in the plots in Figs. \ref{fig_A30-B3}, \ref{fig_A60-B1} \&
\ref{fig_A90-B1}. The variation of $\rho$ on the leading side is more
or less steady, while on the trailing side it often varies very
rapidly due to the aforesaid reasons.

The trend is that the asymmetry seen in $\rho$ will be higher for
larger values of $\alpha$ and smaller values of $S_{\rm L}$ and
hence it is a combined effect of both.  As mentioned earlier, the
rotation effects are more at higher $\alpha$ and hence the larger
asymmetry. Since the range of emission altitudes covered by the
emission spots for inner field lines are higher than that of the
outer field lines, the corresponding values of $\rho$ also will have
a higher range and a higher asymmetry than the ones for the outer
field lines. However, there are variations in the amount of
asymmetry with in the field lines when the $\alpha$ varies, and
hence a steady variation of asymmetry with $\alpha$ will not be
observed as for the emission heights. 

Comparing the $\rho/\rho_{\rm L}$ plots for figs. \ref{fig_A60-B1} 
\& \ref{fig_A90-B1}, we can find on the leading
side that the radius of curvature gets reduced when $\alpha$
increases from $60^{\circ}$ to $90^{\circ}$ for all field lines. But
on the trailing side the behaviour is slightly different.  The inner
field lines ($S_{\rm L}=0.1 ~\&~ 0.3 $) have the radius of curvature
slightly reduced on the trailing side, while an increment in radius of
curvature is seen for outer field lines ($S_{\rm L}=0.1 ~\&~ 0.3 $),
when $\alpha$ increases from $60^{\circ}$ to $90^{\circ}.$  Another
comparison of the $\rho/\rho_{\rm L}$ plots for the
figs. ~\ref{fig_A30-B1} ~\&~ \ref{fig_A60-B1} obviously shows the same
trend for the leading side.  However, the variation of $\rho/\rho_{\rm L}$ 
on trailing side shows a slightly different behaviour, varying
among field lines.  The variation of $\rho $ on the trailing side is
not in a steady pattern as on the leading side. The competing
curvatures due to rotation induced curvature and the intrinsic
field line curvature gives a highly varying pattern for $\rho $ on the
trailing side.

\subsubsection{Spectral Intensity, $I_{\omega}$}
 
The derived spectral intensity curves reflect the asymmetry inherited 
from the variation of $\rho$ with pulse phase, combined with the effects 
of the value of $\gamma$.  
In particular, it is readily seen that the intensity dramatically  
evolves with $\gamma.$  For lower values of $\gamma$, the $I_{\omega}$ 
has a stronger leading part while for higher values of $\gamma$ the 
$I_{\omega}$ has a stronger trailing part.
This effect can be better understood by considering the parameter
$\rho_{\rm p}$ (defined in Eq.~(\ref{eq_rho_p})), which gives the value 
of $\rho$ at which the spectral intensity peaks, for given values of 
frequency and $\gamma$.  For values of $\rho$ greater than or less than 
$\rho_{\rm p}$, the spectral intensity falls monotonically.  
Further, the peak value of the spectral intensity also depends on the 
specific value of gamma, as per Eq.~(\ref{eq_spec_Int}).  Hence, the 
variation of spectral intensity with pulse longitude can be inferred 
from that of $\rho$ with longitude, for different values of gamma.  This 
is illustrated in fig. ~\ref{fig_A30-B1-SPEC} where $\rho/\rho_{\rm p}$ 
and the corresponding $I_{\omega}(\phi')$ are plotted side by side as a 
function of the rotation phase, for specific combinations of parameters.  
    
Three different cases are useful to consider.
For situations where $ \rho/\rho_{\rm p}$ is greater than 1.0 for the 
entire pulse window, the spectral intensity curve shows a maximum at 
$\phi'=0$ and falls asymmetrically on either side, with the reduction in
intensity being larger for the trailing side, due to the faster increase
of $\rho$.  This effect, which is seen for relatively small values of 
$\gamma$ (less than 400-600), naturally leads to asymmetric pulse profiles, 
with possibilities for sharply one-sided profiles.  It is 
interesting to note that in some cases, the intensity on the trailing
side can drop to negligible values, compared to its value at the 
corresponding longitude on the leading side.  This could be a natural
explanation for the one-sided cones reported in literature, and is 
discussed in more detail in \S\ref{sec:partial-cones}.

For situations where $ \rho/\rho_{\rm p}$ is less than 1.0 for the
entire pulse window, the spectral intensity curve shows a minimum at
$\phi'=0$ and rises asymmetrically on either side, with the increase 
being larger on the trailing side.  However, the contrast in the 
intensity levels between leading and trailing sides is typically not 
as high as for the first kind.  This behaviour is seen for relatively 
large values of $\gamma$.
  
For intermediate cases, where values of $\rho/\rho_{\rm p}$ less and 
greater than 1.0 can occur at different points in the pulse longitude 
window, we see more complicated shapes for the spectral intensity 
curves, including multiple maxima at different pulse longitudes.  

For a given geometry and $s_{\rm L}$ value, the transition through
these 3 different cases can take place as $\gamma$ is varied over a
range of values.  Thus, lower values of $\gamma$ tend to produce
profiles with strong leading and weak trailing intensities which get
converted to weak leading and strong trailing kind as the $\gamma$
increases to very high values (e.g. 2nd and 3rd panels from the top in
fig. \ref{fig_A30-B1-SPEC}).  Though the contrast in intensity is less
for the latter, the absolute value of the spectral intensity is
higher, due to the $\gamma$ dependence in Eq.~(\ref{eq_spec_Int}).

Most of the asymmetric intensity profile effects become more dramatic 
for inner field lines and for more orthogonal rotators (larger $\alpha$) 
and smaller values of $\beta.$

\subsubsection{$\theta$ and $\phi$}

The values of $\theta$ and $\phi$ are asymmetric on the leading and
the trailing sides of the profiles, while they are symmetric in the
non-rotating case (G04).  This asymmetry is also an effect of
rotation.  Since $\theta$ and $\phi$ are functions of $\phi'$ 
their values are affected by the aberration phase shift which is 
different on leading and the trailing sides.  Their values are dominantly 
decided by $\alpha$ and $\beta$.  As expected, the shape of the $\phi$ 
curve closely resembles the S-shape of the polarization angle curve
 (see fig.~\ref{fig_A30-B1}).
\subsubsection{Mis-alignment angle, $\eta_{\rm mis}$}       

The Mis-alignment angle $\eta_{\rm mis},$ defined as $\eta_{\rm
  mis}=\cos^{-1}({\hat \textbf n}\cdot{\hat \textbf v}),$ gives an
estimate of the offset between $\hat \textbf n$ and the estimated
$\hat \textbf v.$ In principle, for a perfect estimation of the
emission point, the line of sight should exactly coincide with the
velocity vector, and hence $\eta_{\rm mis}$ should be zero. However,
in actual computations, $\eta_{\rm mis}$ always has a small, finite
value.  A quick check of the accuracy of the computation is provided
by the value of $\eta_{\rm mis}$ : a lower value implies a higher
precision of estimation of the emission spot, and vice versa for a
higher value.  A rough classification that can be taken for the
precision of the estimation is: a value of $\eta_{\rm mis} \ll 1$
indicates a highly precise estimation of the emission spot, and vice
versa for $\eta_{\rm mis} \gg 1.$ By this scheme, we find that there
is satisfactory precision for all estimations for $r$ within 20 \% of
$r_{\rm L}$ (see fig.~\ref{fig_MISANG-INNER}). In some cases the
$\eta_{\rm mis}$ exceeds 1, but only when $r/r_{\rm L}> 0.2.$ However,
according to established observational results radio emission heights
are restricted within 10 \% of $r_{\rm L}$ for normal pulsars
(e.g. Kijak 2001), and hence our method is quite satisfactory in this
regime of interest.

\subsection{Effects of Parameters}

The above described behaviour of the height of emission spots, radius 
of curvature and spectral intensity are strongly dependent on the 
parameters like geometry, field line location, radio frequency and 
Lorentz factor of the particles.  In some cases, there is a complex 
interplay between the dependencies on these different parameters.  
Here, we explore some of these effects in detail.
 
The generic effects of $\gamma,$ $\alpha$ and $\beta,$ $\omega$ and 
$S_{\rm L}$ are listed briefly below in an order that may characterize 
the hierarchy of their effects on total intensity profiles.  

\subsubsection{Inclination angle, $\alpha$} \label{alpha-effect}
The parameter $\alpha $ is a major driver of the effects of rotation, 
and has the strongest influence on our results and conclusions.  The
rotation effects (leading-trailing asymmetry of $r,$ $\rho$ and
$I_{\rm \omega}$) are more prominent for large values of $\alpha$ and 
less for small values of $\alpha.$  The range of emission altitudes 
is found to be relatively high for lower values of $\alpha$, and 
relatively low for higher values of $\alpha,$ being the lowest for 
$\alpha=90^{\circ}.$  Like wise, $\rho$ appears to reach higher values 
for higher $\alpha$ and vice versa for lower $\alpha.$

\subsubsection{Normalized foot value of the field lines, $ S_{\rm L} $} 
The effect of moving from inner to outer regions of the magnetosphere
(increasing $S_{\rm L}$ values) also has a very dramatic effect on
the results.  Rotation effects are strongest for the innermost field
lines, and decrease significantly for larger $S_{\rm L}$ values.  For
relatively small values of $S_{\rm L}$ (usually $\le$ 0.3), the leading
part has emission heights that vary relatively gently with $\phi'$, 
whereas the trailing part shows steeply rising emission height.  The
emission heights become less asymmetrical for increasing values of
$S_{\rm L}.$  The values of $\rho$ steadily increase with decreasing
$S_{\rm L}$ (i.e. inner field lines) on the trailing side.  Dramatic
effects such as very large, peaked values of $\rho$ on the trailing
side, owing to the mutual cancellation of intrinsic and rotation
induced curvatures, are seen only on inner field lines.  This peak
shifts closer towards zero pulse phase as the value of $ S_{\rm L}$ 
becomes smaller.  For outer field lines, the profiles are much more 
symmetric, and since $\rho/\rho_{\rm p} < 1$ for a significant range 
of pulse phase on either side of $\phi'=0$, the profiles more often
exhibit minima at $\phi'=0$, even for moderate values of $\gamma$.
 
\subsubsection{The lorentz factor, $\gamma$} \label{gamma-effect}
The effect of $\gamma$ has a very clear signature on the asymmetry 
of the spectral intensity profiles.  For lower values of $\gamma$, 
there can be strong asymmetries with leading side stronger 
than the trailing side, and maxima at $\phi'=0$.  For larger values 
of $\gamma$, the sense of this asymmetry can reverse, with trailing 
side becoming stronger than the leading, and a minima at $\phi'=0$ ; 
however, the degree of the asymmetry, as measured by the ratio of 
the intensities at corresponding longitudes, is generally less than 
that for the case for the low $\gamma$ values.

\subsubsection{Emission frequency, $\omega$}
The frequency of emission, $\omega$, acts as a counter to the effect 
of $\gamma$, though in a relatively weak manner, as can be 
understood from Eqs. ~(\ref{eq_spec_Int}) and ~(\ref{eq_rho_p}).
Thus, an increase in $\omega$ produces changes which can be 
compensated by a corresponding change in $\gamma$ by a factor
proportional to $\omega^{1/3}$. In certain cases, this could
result in profiles where the sense of asymmetry between leading and
trailing sides could reverse over a large enough range of radio
frequencies.  Such effects are seen sometimes in some real profiles.

\subsection{Realistic profiles}

One of the significant results from our simulation studies is that
the possible regions of emission associated with a given annular
ring of field lines (characterised by a constant value of $S_{\rm L}$) 
are visible over a wide range of pulse phase, albeit with
different intensity levels. This aspect, combined with the results
for field lines with different values of $S_{\rm L}$, leads to the
conclusion that a very large fraction of the pulsar magnetosphere
is potentially visible to us.  This results in simulated pulsar
profiles that are very different from the observed profiles of
real pulsars which appear to have well defined emission components,
restricted in pulse phase extent to occupy only some fraction of
the on-pulse window.  This disparity with the observed profiles
persists even after we incorporate into our simulations the models
of discrete, annular conal rings of emission.  

Hence, in order to reproduce realistic profiles matching with
observations, we need some additional constraints for the emission
regions.  In the most general case, such non-uniformities in the
distribution of emission regions can exist in any of the three
coordinate directions, viz.  $r$, $\theta$ and $\phi$.  Non-uniformity
of emission in the $\theta$ direction is achieved in some sense at the
basic level, by considering only discrete sets of $S_{\rm L}$ values
for active emission regions.  As discussed, this is not enough to
constrain the intensity variations to reproduce realistic pulsar
profiles.
 
The possibility of non-uniform emission along the $\phi$
coordinate could help produce discrete emission components in the
observed profiles.  As seen in fig. ~\ref{fig_A30-B1}, for a given
value of $S_{\rm L},$ the emission at different pulse longitudes
originates at widely different $\phi$ locations.  If the sources of
charged particles were located only at fixed $\phi$ points along the
ring of constant $S_{\rm L},$ then these could be arranged to modulate
the simulated intensity pattern with a suitable ``window'' function,
to obtain discrete emission components in the observed profile.
However, the well known phenomenon of sub-pulse drift argues against
this being a viable option.  Sub-pulse drift, which is now believed to
be fairly common in known pulsars (e.g. Velterwede et~al. 2006),
wipes out any azimuthal discretization of sources of emission in the
pulsar magnetosphere -- it would lead to ``filling up'' of the intensity
average profile over a given range of pulse longitude, as is observed
in drifting sub-pulses that occur under any discrete emission component
in known pulsars.

The third option is to have non-uniform emission in the radial 
direction, i.e. preferred heights of emission for a given set of field 
lines.  Since the contributions at different pulse phases are from 
different heights, this would naturally lead to non-uniform 
distribution of intensity in pulse phase, resulting in realistic
looking pulse profiles.  The idea of preferred heights of emission
in the pulsar magnetosphere is not entirely new -- the ``radius to
frequency mapping'' model for pulsar emission postulates different
heights for different frequencies, with the height of emission
increasing for decreasing frequency values (Kijak \& Gil 1997,1998; 
GG01, GG03).
    
\begin{figure}         
\begin{center}     
\epsfxsize= 6.5 cm  
\epsfysize=0cm
\rotatebox{0}{\epsfbox{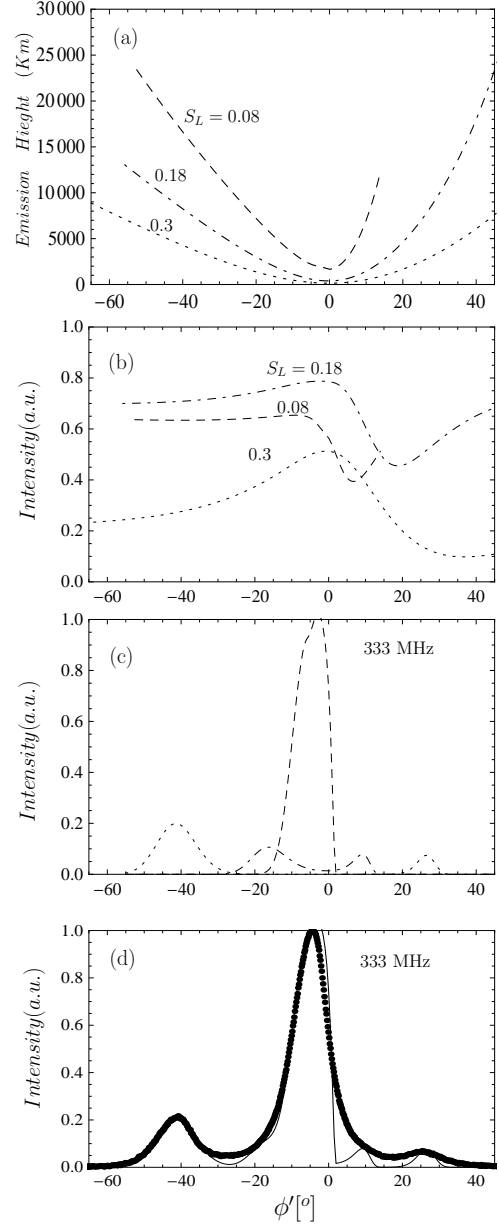}}
\caption[short_title]{\small PSR B2111+46 at 333 MHz : The emission 
height for each $S_{\rm L}$ value associated with a particular 
component is plotted in panel (a). The simulated and un-modulated 
spectral intensity curve corresponding to each $S_{\rm L}$ value is 
plotted in panel (b).  The simulated sub-components, after applying 
Eq.(\ref{Eq-Mod-spec}) for the best fit values given in 
Table.\ref{tabsim} are shown in panel (c).  The sum-total of the
simulated sub-components giving the final profile (solid curve)
is shown superposed with the observed profile (dotted curve) in 
panel (d).
\label{2111+46-333}}
\end{center}  
\end{figure}     

Preferred emission heights of emission with a spread in the $r$
direction, can be modeled as a multiplication of the spectral intensity
${I}_{\omega}(\phi')$ with a modulating function, $F^{\rm i}_{\omega}
(\phi')$.  For modelling a profile as a sum total of emissions from a 
core and several discrete conal regions, the modulated spectral intensity
can be expressed as
\begin{eqnarray}\label{eq_mod_funct}  
{I}_{\omega}(\phi') &=& \sum_{i}^{N} {I}^{\rm i}_{\omega} (\phi')\,\,
F^{\rm i}_{\omega} (\phi')~,   \label{Eq-Tot-spec} \\
F^{\rm i}_{\omega} (\phi')&=& A^{\rm i}_{\omega} \exp\left[-\left(
\frac{r^{\rm i}_{0}(\phi')- H^{\rm i}_{\omega }} 
{2\, \Delta H^{\rm i}_{\omega}}\right)^2 \right]~,  \label{Eq-Mod-spec} 
\end{eqnarray}          
where the index $i$ represents, a corresponding pair of
leading-trailing components presumed to be arising from a particular
ring of field lines; while $i=1$ exclusively represents the central
core component. Here $N$ is the total number of such discrete emission
components (for example, $N=3$ would correspond to a 5 component
profile forming a central core and two pairs of conal components),
${I}_{\omega}(\phi')$ is the total spectral intensity from all field
lines combined, while ${I}^{\rm i}_{\omega} (\phi')$ is the spectral
intensity from the $i$th ring of field lines.  For a given emission
region, $H^{\rm i}_{\omega}$ represents the mean height, and $\Delta
H^{\rm i}_{\omega}$ represents the spread of the region. The variable $r^{\rm
  i}_{0}(\phi')$ represents the values of emission altitude for each
value of $\phi'$ estimated by the simulation method as described
earlier, for the $i$th ring of field lines.
   
To map this intensity as a function of rotation phase as seen
in the observer's frame, the effects of retardation and aberration
need to be included explicitly. The effect of aberration is estimated
by default and the $\phi'$ is inclusive of the aberration phase shift.
The retardation phase shift is to be estimated from the value of
$r$ corresponding to the emission spot. The rotation phase
corresponding to ${I}_{\omega}$ is updated after adding the
retardation phase shift $d\phi'_{\rm ret}$ with $ \phi',$ and this
is represented by the mapping of the ordered pair
$(\phi',{I}_{\omega})\rightarrow (\phi'+d\phi'_{\rm ret},{I}_{\omega}).$ 
The $d\phi'_{\rm ret}$ can be estimated as (G05) :
$$d\phi'_{\rm ret}=\frac{1}{r_{\rm L}}({\vec r}\cdot{\hat n}),$$
where ${\vec r}=r{\hat e}_{\rm r}$ and the expression for 
${\hat e}_{\rm r}$ is given in \S\ref{app:approx}.  The height of 
emission ($ H^{\rm i}_{\omega }$) and the normalized foot value 
($S^{\rm i}_{\rm L}$) corresponds to the peak of the $i$th component 
of the profile.  The $\Delta H^{\rm i}_{ \omega}$ and 
$A^{\rm i}_{\omega}$ are model parameters.

\begin{table}
\centering   
\begin{minipage}{100mm}
\caption{The parameters for simulating the  profiles of PSR B2111+46 }\label{tabsim}
\begin{tabular}{ccccccccccccccccccccccccccccccc}
\hline \\ 
\multicolumn{1}{c}{Frequency} & ${\rm i}^{a}$
& \multicolumn{1}{c}{$H^{i}_{ \omega}$}
& \multicolumn{1}{c}{$\Delta H^{i}_{ \omega}$}
& \multicolumn{1}{c}{$A^{i}_{\omega}$ }
& \multicolumn{1}{c}{$S_{\rm L}$ }
& \multicolumn{1}{c}{Lorentz factor}  \\

\multicolumn{1}{c}{MHz} &     
&\multicolumn{1}{c}{Km}    
&\multicolumn{1}{c}{Km}   
& \multicolumn{1}{c}{ }         
& \multicolumn{1}{c}{ }
& \multicolumn{1}{c}{$\gamma$}
\\
\hline \hline
  
333 & 1 &  1500 & 600 & 1.8  & 0.08 & 750 \\
    & 2 & 1834 & 500 & 0.14 & 0.18  & 750 \\
    & 3 & 3800 & 500 & 0.7  & 0.3 & 500 \\ 

408 & 1 &  300 & 850 & 2.2 & 0.09 & 750 \\
    & 2 & 1200 & 750 & 0.1 & 0.22 & 750 \\
    & 3 & 3000 &  400 & 0.9 & 0.35 & 500 \\ 

610 & 1 &  200 & 700 & 1.7 & 0.13 & 750 \\
    & 2 &  500 &  450 & 0.3 & 0.31 & 700 \\
    & 3 & 2600 & 500 & 1.5 & 0.35 & 550 \\ 
\hline 
\end{tabular}\\
\\ \small{$^a$ $i=1$ represents the core component, \\
$i=2$ represents the inner conal component,  \\
$i=3$ represents the outer conal component.}
\end{minipage}
\end{table}

\begin{figure*}         
\begin{center}
\epsfxsize= 14 cm
\epsfysize=0cm
\rotatebox{0}{\epsfbox{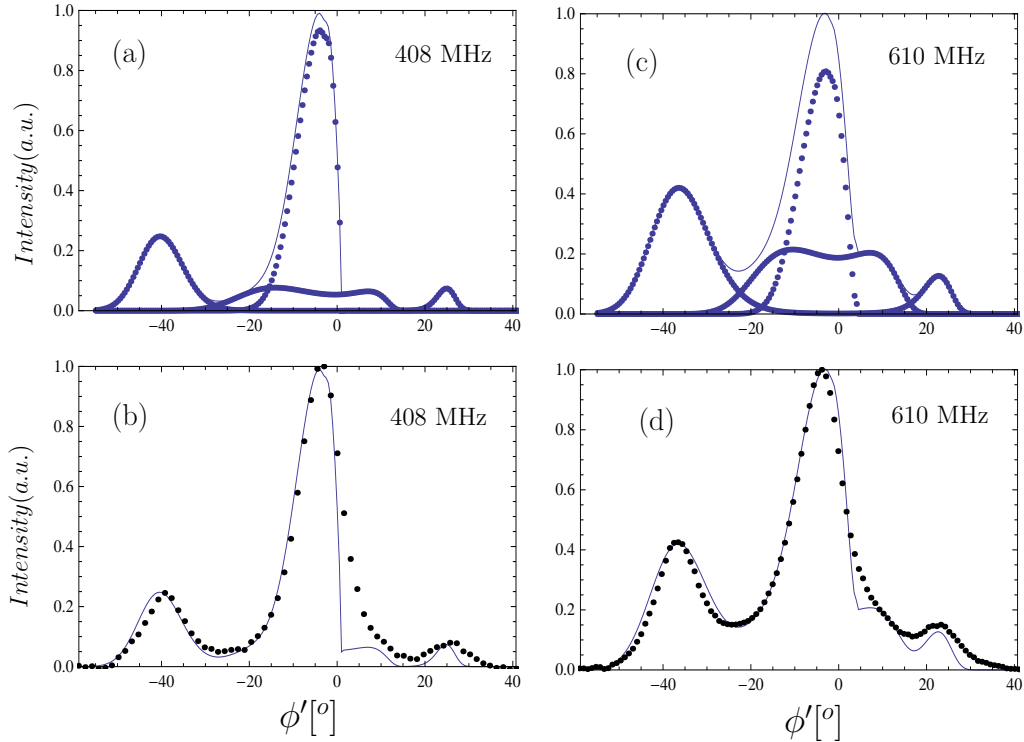}}
\caption[short_title]{\small PSR B2111+46 at 408 \&  610  MHz:
In panels (a) and  (c) , the dotted curves shows the simulated 
sub-components and the continuous curve shows their sum-total.
In panels (b) and  (d), this final simulated profile (solid curve) 
is superposed with the observed profile (dotted curve).  
\label{2111+46-ALL}}
\end{center}        
\end{figure*}      

\subsection{Profiles for PSR B2111+46 : a test case}

Using the afore said methods for simulation of pulse profiles, we have
attempted to reproduce the intensity profiles of PSR B2111+46 obtained
from EPN data base and GMRT data, at multiple radio frequencies. This pulsar has a
multi-component profile, with a well identified core component and 2
cones of emission (e.g. Zhang et~al. 2007).  It has a rotation period
of 1.014 sec and $\alpha=14$ and $\beta=-1.4$ (Mitra \& Li 2004). The
other parameters used in the simulation are listed in Table
\ref{tabsim}.  The values of emission heights $H^{\rm i}_{\omega}$ and
field line locations $S^{\rm i}_{\rm L}$ for the discrete emission
components are the values from  estimates employing  the method
given in Thomas and Gangadhara (2009).  The zero phase of the profile
is fixed on the basis of the analysis of core emission of this pulsar,
using the method developed in the the same work.

The simulation method is illustrated in fig.~\ref{2111+46-333}. 
Values of $S_{\rm L}$ corresponding to the core and conal components 
are employed in generating the emission height plots in panel (a) of 
fig.~\ref{2111+46-333}. The corresponding spectral intensity plot for
each of these $S_{\rm L}$ values, for the final best fit choice of 
$\gamma$ (in Table \ref{tabsim}) is shown in panel (b) of this figure. 
The individual components generated after applying the best fit height
function are shown in panel (c) and the sum total intensity profile 
is shown in panel (d), along with the observed profile.  
Best fits of these profiles to the observed data were obtained by 
varying $ A^{\rm i}_{\omega}$ and $\Delta H^{\rm i}_{\omega}$ in
the function $F^{\rm i}_{\omega}(\phi')$, and by varying the value
of $\gamma$ in the range 100 to 1000.  The same procedure is repeated 
for 408 MHz and  610 MHz  profiles and the results are shown in 
fig.~\ref{2111+46-ALL}.  All the final parameters and best-fit results 
are summarised in Table.\ref{tabsim}.  

An encouraging first order match between the simulated and observed
profiles has been achieved (see panel (d) of fig.~\ref{2111+46-333} \&
panels (b) and (d) of fig.~\ref{2111+46-ALL}).  The core component is
quite well fit for most of the cases, and so are the leading conal
components.  There is some mismatch in the widths of the conal
components, especially for the trailing side, where the real data
shows a smoother blending of the components, compared to the simulated
profile where the components appear more narrow and relatively well
separated.  It is remarkable that with a single value of $A^{\rm
  i}_{\omega}$ for a leading-trailing pair of cones of emission, the
ratios of the peak values of the intensity of the leading and trailing
components of the cones match so well with the real data.  It is also
interesting to note that the best fit values for $\gamma$ are very
similar for a given emission component, at different frequencies,
supporting a model of a common bunch of accelerated charged particles
being responsible for the emission at different frequencies.  Further,
that the spread of $\gamma$ values across the different emission
components is also quite small, indicates very similar operating
conditions over most of the magnetosphere.  The best fit values for
$\Delta H^{\rm i}_{\omega}$, though reasonable, are somewhat large in
amplitude, indicating somewhat extended emission regions in the
magnetosphere.

We note that these relatively large values of $\Delta H^{\rm i}_{\omega}$
and some of the limitations of the fits may be due to the lack of 
some generalizations in our model.  These include factors like 
coherency of emission, a realistic spread of $\gamma$ values around 
the mean values obtained here, as well as a realistic spread in the
values of $S_{\rm L}$ due to finite thickness of the rings of emission
on the polar cap.  Whereas a detailed treatment of all of these is
beyond the scope of this work and will be taken up later, some basic 
inferences can still be drawn.  For example, if a small range of 
$S_{\rm L}$ values around the mean is considered, it is easy to 
argue that much of the width of a profile component can be filled up
by radiation from such a bunch of field lines.  This can be understood
from panel (a) of fig. ~\ref {2111+46-333}, where a line of constant
height intersects the curve for a given field line at two points, one
each on the leading and trailing side.  The phase of this point of
intersection will move systematically as we go to neighbouring field
lines.  Hence, wider profile components can be achieved with smaller
values of $\Delta H^{\rm i}_{\omega}$.  Furthermore, due to the 
asymmetry in the emission height curves, the shift in phase with 
change of $S_{\rm_L}$ is more on the trailing side, which would 
naturally lead to broader component widths and better `blending' of 
the components in the profile, something that is not as easily 
achievable by having a large range of emission heights (as the 
shift of phase for a given separation of heights on a given field
line is lesser on the trailing side).  One indicator of the significance
of the spread of $S_{\rm_L}$ values is the amount by which $S_{\rm_L}$
needs to be changed to move the peak of one conal component to the 
point half-way to the peak of the next conal component.  Not very
surprisingly, our rough estimates show that the required change 
in $S_{\rm_L}$ is close to the half-way point to the $S_{\rm_L}$ 
value of the next cone, which would indicate a closely packed
structure of concentric rings.  
    
The component profiles could be further influenced by considering a
distribution of $\gamma$ values associated with the emitting particles.
We have found that significant shifts in the peaks of the leading and 
trailing pair of components for simulated profiles are obtained for 
lower $\gamma$ values ($\gamma < 500$), while the peak positions appear 
almost frozen for increasing $\gamma$ values. Thus it is realistic to
assume that a spread of $\gamma$ values can broaden the emission
components. This factor also may reduce the $\Delta H^{\rm i}_{\omega}$ 
required to effect a good fit.    
        
Nevertheless, we would like to point out that there is only one unique
combination of the parameters that can produce a profile which is
similar to the observed one. We have not found any degenerate
combination of values for the parameters that are shown in
Table.\ref{tabsim}.  Thus, the similarity of the simulated profiles
with the observed ones gives an assurance that, we should be able to
simulate the observed profiles with greater similitude with a model
overcoming the above-said limitations.

\subsection{Core emission}

The generation of the profile components for PSR B2111+46 described
above naturally leads to a discussion on the core emission. In fact,
the study of the phenomena of core emission has spawned enormous amount
of literature.  Perhaps the most notable ones are the landmark
work by Rankin (1983) that systematized the pulsar emission profile
into `core' and `cone', and the succeeding works by Rankin (1993a
\& 1993b) that further developed the core-cone classification
scheme. The hollow cone model was invoked to explain the geometry
(e.g. Taylor \& Stinebring 1986) and the origin of core emission.
Radhakrisnan and Rankin (1990) have conjectured that the emission
mechanism for cores might be different from that of cones, owing to
the behaviour of polarization position angle curve near the core
being different from the rotating vector model.  However, there are
no satisfactory theoretical grounds for postulating diverse mechanisms 
for cores and cones.  A major difficulty that curvature radiation 
encounters in explaining the core emission is the insufficient curvature 
of the almost straight field lines in a region relatively close to the 
pulsar polar cap. Since the intensity of emission is proportional to 
$1/\rho^2$, the values of $\rho$ provided by the intrinsic curvature 
of the field lines is too large and hence insufficient to generate 
enough intensity of emission typically observed for core component.  
This factor even prompted invoking other emission mechanisms for
explaining core emission (e.g. Wang et~al. 1989).

In our simulation studies, the presence of the core component comes
about quite naturally.  It can be seen from all the plots of spectral 
intensity in \S\ref{app:figs} that the emission from regions near 
the profile centre is comparable (for higher $\gamma $ values) or 
even somewhat higher (for lower $\gamma $ values) than that from 
regions in the wings of the profile.  This happens because we get
low values of $\rho$ for inner field lines near $\phi' \approx 0$, 
which are comparable to that of outer field lines, and this occurs 
consistently for all the combinations of $\alpha$ and $\beta$ (see 
the panels for $\rho/r_{\rm L}$ in \S\ref{app:figs}).  The reason 
is that rotation induces significant curvature into the trajectory 
of particles, even though they are confined to move along the nearly 
straight inner field lines.
 
The forces of constraint act  in such a way that  the particle 
is hardly allowed to deviate away from the field line on which it is moving,
 and the resulting scenario is discussed in detail in TG07. 
Due to the co-rotation of the field lines 
and the action of the aforesaid  forces of constraint 
the charged particles are added with a  velocity component in the 
direction of rotation, which is nearly perpendicular to the velocity component parallel to the  field line, in the observer's frame. This induces an additional  curvature in the trajectory of the particle  
and  makes it significantly different from that of the field line curvature in the observer's frame of reference (TG07). Hence the trajectory of charge particles moving on almost straight field lines 
 near the magntic axis can have a highly curved trajectory and hence  a relatively low 
 value of radius of curvature that is  significantly different from that of the field lines. 
 This scenario  allows for significant 
emission near the central region of the profiles.  By applying 
Eq.~(\ref{Eq-Tot-spec}) and Eq.~(\ref{Eq-Mod-spec}) appropriately, 
as described earlier, profile shapes resembling strong core components 
can be easily generated.  Hence by applying our method, we provide 
a natural explanation for core emission, that circumvents the issue 
of too high $\rho$ that precludes a strong core component with 
curvature emission.

In the simulation of the profiles for PSR B2111+46, we find that the
core originates from have relatively inner field lines and lower 
emission heights than the cones.  Assuming the same mechanism of 
emission, viz. curvature radiation, for the core and conal component,
we are able to produce a simulated core component that matches quite 
well with the observed profile.  We notice that the best-fit values 
for the amplification factor $A^{i}_{\omega}$ found in the simulation 
for the core component (Table ~\ref{tabsim}) are comparable to those
of the cones.  These values are not unduly high, considering the 
situation that the density of plasma should be relatively higher for 
the lower altitude and hence an additional factor for relatively stronger
emission at lower altitudes.  Our profile-matching of PSR B2111+46
thus shows that strong core emission can originate from inner field
lines due to curvature emission.

\subsection{Partial cones}\label{sec:partial-cones}

According to LM88, partial cone profiles are the ones in which one side
of a double component conal profile is either missing or significantly 
suppressed.  These are recognised by the characteristic that the steepest 
gradient of the polarization position angle is observed towards one edge 
of the total intensity profile, instead of being located more centrally 
in the profile as the rotating vector model postulates.
LM88 speculated that this happens when the polar cap is only partially 
(and asymmetrically) active. It is significant that, out of the 32 
pulsars listed in LM88 that display the partial cone phenomena, as many
as 22 have the steepest gradient point occurring in the trailing part 
of the profile.  In other words, most of the partial cone profiles
show a strong leading component and an almost absent trailing component.

\begin{table*}
 \centering
 \begin{minipage}{100mm}
  \caption{The parameter values employed  in 
fig.\ref{PARTIAL} }\label{tabpar}
  \begin{tabular}{cccccccccccccccccccccccccc}
  \hline \\ 
   \multicolumn{1}{c}{Panel} &
 $\alpha$
& \multicolumn{1}{c}{ $\beta$} 
 & \multicolumn{1}{c}{$S_{\rm L}$ }
&\multicolumn{1}{c}{$H^{i}_{ \omega}$}
&\multicolumn{1}{c}{$\Delta H^{i}_{ \omega}$}
\\

\multicolumn{1}{c}{No.}  
&\multicolumn{1}{c}{$[^o]$}
&\multicolumn{1}{c}{$[^o]$}  
&\multicolumn{1}{c}{ }   
 & \multicolumn{1}{c}{$Km$ }
 & \multicolumn{1}{c}{$Km$}    
\\
\hline \hline

1 & 90 & 1 & 0.3 & 1000 & 200 \\ \\
2 & 90 & 1 & 0.5 & 1200 & 150  \\ \\
3 & 60 & 1 & 0.3 & 1000 & 200 \\ \\
4 & 60 & 1 &  0.5   &   1400 & 150 \\ \\
5 & 30 & 1 &  0.1   &   2500 & 500 \\ \\
6 & 30 & 1 &  0.3   &   1000 & 150 \\ 
\hline 
\end{tabular}
\end{minipage}
\end{table*}

There are two possible scenarios that have been postulated to explain
partial cones : (1) only a part of the polar cap is active (this works
for both kinds of partial cones) or (2) the A/R effects are so large 
as to shift the entire active region of the intensity profile towards 
the leading side (this works for the strong leading type partial cones,
which are the majority).  However, Mitra et~al (2007) studied several 
pulsars with partial cones with very high sensitivity observations and
found that the almost-absent parts of the cones do flare up occasionally
and show emission for about a few percentage of the total time.  This
tends to rule out both the scenarios above, and requires an explanation 
where the intensity is naturally suppressed in one side of the cone.

In our simulation studies, one sided cones appear as a natural by-product.
We notice that for smaller values of $\alpha,$ inner field lines and 
lower $\gamma$ values, the intensity profile is almost always significantly
suppressed on the trailing side, as compared to the leading side. The 
reason for this is quite obvious. As explained earlier, the $\rho$ for
inner field lines is highly asymmetrical between the leading and trailing 
sides -- it remains more or less steady on the leading side, while on the 
trailing side it shoots up to a high value and then falls.  Whenever 
the $\rho$ shoots up such that  $\rho/\rho_{\rm p}\gg 1$, the spectral 
intensity is significantly reduced.  On the other hand, on the leading 
side we mostly have $\rho/\rho_{\rm p}\approx 1$ and hence the 
spectral intensity is significant there.  For relatively lower values 
of $\gamma,$ $\rho_{\rm p}$ reduces and hence there is a greater chance 
of having $\rho/\rho_{\rm p}\gg 1$, while for higher values of $\gamma$, 
$\rho/\rho_{\rm p}$ drops down and eventually becomes closer to 1. 
Hence the intensity plots shown in \S\ref{app:figs} are stronger on the 
leading side at low $\gamma$, and stronger on the trailing side at 
high $\gamma$.  However, it is to be noted that (i) the values of $\gamma$
required to achieve stronger trailing side profiles are very high  -- 
usually significantly more than 1000; whereas, for more typical values
of $\gamma$, we get the stronger leading side profiles and (ii) the 
intensity contrast obtained for the stronger leading side profiles is 
much larger and striking, compared to that for the stronger trailing side
profiles.  Both these facts argue naturally for a strong preponderance
of one sided cones with stronger leading side profiles, as is statistically
seen in the results of LM88.

To further illustrate the idea, we have generated profiles as shown in 
fig. ~\ref{PARTIAL} that resemble partial cone profiles by using our 
simulation technique for specific combinations of parameters, which are
listed in Table ~\ref{tabpar}.  The thin line curve represents the
un-modulated profile, which is simulated by assuming that the emission
is uniform all along the field line, while the thick line shows the final
modulated profile.  The active region of the final profile is clearly 
shifted to the leading side, due to the afore said behaviour of $\rho.$ 
The suppression of the intensity on the trailing part in comparison to 
the leading side is seen in all the plots and is most dramatic for the 
inner field lines, for low values of $\gamma$, and for large values of
$\alpha$.

\subsection{Studying the mechanisms of emission : future prospects}   

In this section, discuss some of the future possibilities from the
present work.  Though we have developed a model under certain specific
conditions and demonstrated some useful results from the same, it
has significant potential for applicability under diverse circumstances
and conditions.  The $r$ and $\rho$ of the emission spot are two of the
fundamental ingredients for computing the intensity of emission within 
any model of radio emission for pulsars.  These values, along with other
parameters that we have calculated in our method after explicitly 
taking into account of effects of rotation and geometry, are applicable 
for any model of radiation that precepts the condition embodied in
Eq.(\ref{eq_los_vel}), i.e. having the radiation beam aligned with
the velocity vector and line-of-sight. Thus the present method of
computing $r$ and $\rho$ is well suited for studying curvature radiation 
models in vacuum approximation. The profiles simulated by these models 
can be compared with the observed ones to check their veracity. 

Though we have employed single particle curvature radiation formulation,
it is well known that this cannot explain the extremely high luminosities 
seen in  typical pulsar radio emission.  Coherent emission from bunches 
of charged particles have been argued to be necessary for explaining the 
high luminosities (eg. Ginzburg et~al. 1969, Melrose 92, Melrose 2006). 
The model of coherent emission constructed by Buschauer and Benford (1976) 
considered relativistic charge and current perturbations propagating 
through the bunches with N number of charges, which boosted the emission 
much above typical $ N^2$ factor.  Further, they have shown that the 
characteristic frequency will  be significantly shifted to higher values 
than the typical $\approx 1.5 \gamma^3 c/\rho.$  However, the extremely  
short lifetime of  these moving sheets of plasma (bunches) made it 
implausible to radiate, and due to this reason these
 emission models were almost forgotten. In later years, 
the possibility of formation of Langmuir micro-structures (solitons) due to 
the collective behavior of the plasma brought back the possibility of 
bunched radiation (Asseo 1993). It was shown that the radiation from such 
a bunch could be expressed by just using the classical formula for curvature 
radiation (Asseo 1993).  Melikidze et~al. (2000) considered the three 
component structure of charge distribution for solitons in the pulsar
magnetosphere and obtained a different spectral intensity distribution 
from that of the classical formula for curvature radiation.  However,  
Gil et al. (2004) used single charged bunches of charge Q as equivalent 
to a single particle with the same charge Q, to explore the effects of 
the surrounding plasma on the curvature emission, and showed that sufficient 
luminosity could be produced from curvature emission that matches with the 
observed luminosity of pulsars.

As mentioned in the earlier \S\ref{sec:comp-intensity} our estimates
and results corresponding to altitude, radius of curvature, magnetic
azimuth and magnetic colatitude are equally valid for the case of
coherent and incoherent emission, as long as the vacuum approximation
is invoked. This is because the peak of the emitted beam will be
aligned with the direction of velocity for emission from a source
moving at ultra-relativistic speeds, de-facto in vacuum
approximation. Hence the premise contained in Eq.(\ref{eq_los_vel}) for the
computation of these quantities will remain valid for both of the
cases. For the case of a simple model of coherence for a bunch of net
charge Q, the spectral intensity profile estimated will be similar to
that of the emission from a single particle with charge Q, and
likewise the relative intensity will also be the same.  Hence the
results that we have drawn upon spectral intensity are valid for the
simple case of coherent emission too.  However, invoking models of
coherent radiation with additional features apart from a simple coherent
model, may push the intensity estimates to significantly different
values and the resulting shape of the intensity profile will be
considerably altered.  Two such examples are mentioned in the
following.

 Buschauer and Benford (1976) has shown that both intensity profile
 and characteristic frequency will be altered if the allowance is made
 for the propagation of a charge and current density wave through the
 coherent bunch.Considering this model we find that it can alter the
 shape of the computed spectral intensity curve corresponding to a
 given field line, from that of the present results.  This is mainly
 because of the reason that characteristic frequency $\omega_{\rm c}$
 will be shifted to a higher value than in the case of single particle
 curvature radiation. Another case is the spectral intensity formula
 for emission from solitons having a three component charge structure
 (Eq.(12) in Melikhidze et~al. 2000) which also will yield
 significantly different estiamtes for spectral intensity, from that
 of the spectral intensity estimated for the single particle
 emission. Both of these models are treated in the vacuum
 approximation and hence they satisfy the condition embodied in
 Eq.(\ref{eq_los_vel}), i.e. having the radiation beam aligned with
 the velocity vector and line-of-sight.
This ensures that the method of estimation and hence the results 
corresponding to altitude, radius of curvature, magnetic azimuth,  magnetic
colatitude etc. will be applicable for these two cases also. The  only quantity
that is altered by the inclusion of these models,  from a single particle case,
is the spectral intensity estimate. 
 Nevertheless, these models can  be quite
easily incorporated into our simulation studies, simply by modifying
the form of the spectral intensity expression that is used.

In the emission models where the effects of the surrounding plasma
are considered, the peak of the radiation beam may be offset from the 
velocity vector by a finite angle (Gil et~al. 2004). This requires a
modification to the condition in Eq.(\ref{eq_los_vel}) such that 
${\hat \textbf {n}}\cdot {\hat \textbf {v}} = \eta_{\rm max } $ 
where $ \eta_{\rm max } $ is the value of the angle of offset by which 
the peak of the emission beam is offset from the velocity vector.  
Coupling this with some modifications to our method can deliver the 
values of $r$ and $\rho$ appropriate for this case too.  The analysis 
and results that ensue from all of the above said considerations will 
be discussed in our forthcoming works.\\

 \section{Summary}\label{sec:summary}

We have developed a method to compute the probable locations of
emission regions in a pulsar magnetosphere that will be visible at
different pulse longitudes of the observed profile. 
The effects of geometry and rotation of the pulsar
are accounted in a detailed manner  in this method, which is a very useful new
development.  Our method includes `exact ' and `approximate' techniques
for carrying out the estimation of the relevant emission parameters. 
The `approximate' method is useful for certain extreme regimes of 
parameter space, and for faster computation of the results.  The 
misalignment angle, which provides a good check of the accuracy of the
computations, shows that our method achieves satisfactory precision. 
Besides the exact location of possible emission regions, we are able 
to compute several other useful parameters like the height of emission,
and the radius of the particle trajectory at the emission spot, the 
azimuthal location of the associated field line etc., for different 
combinations of pulsar parameters like $\alpha$ and $\beta.$  Further, 
using the classical curvature radiation as the basic emission mechanism 
(which is apt for a debut level analysis), we are able to compute
the spectral intensity from any emission spot. By assuming a uniform 
emission all along the field lines, we have estimated the spectral 
intensity for a range of pulse phase that the line of sight sweeps through.
We have discussed how realistic looking pulsar profiles can be generated from 
these generalized intensity curves, by assuming specific range of emission
heights along specific rings of field lines.  We have illustrated the 
capabilities of these methods by generating simulated profiles for the 
test case of the pulsar PSR B2111+46, and have shown that fairly good
match with observed profiles can be achieved.  We have also shown how 
 further  detailed (and practical) considerations can help improve this
match. We have shown how our results offer a direct and natural 
explanation for the puzzling phenomena of partial cones that are seen 
in some pulsar profiles.  Our simulations also provide a direct insight
into the generation of the core component of pulsar beams.  
Finally, we have indicated how our method can be extended to incorporate
more sophisticated models for the emission mechanism and produce intensity 
profiles for the same.  These, as well as extension to polarized intensity
profiles, will be taken up as future extensions of the work reported here.

\appendix

\section[]{ Velocity,Acceleration and Radius of curvature of Particle trajectory}\label{app:vel}
\subsection[]{ Expression for velocity and  acceleration}\label{app:vel-accel}
The expressions for velocity $\bf v$, acceleration $\bf a$, radius of
curvature $\rho, $ etc. that are used in the computation described in
\S\ref{sec:exact} and \S\ref{sec:approx} are provided here (see
\S C in TG07 for details).  The
velocity and acceleration (in spherical polar coordinates) of the
charged particle in the laboratory frame can be defined as .
\begin{equation}\label{eq_par_tot}
{\bf v}= \frac{d\,r}{dt}\, {\hat \textbf{
  e}}_r+r\,\frac{d\,\theta'}{dt}\, {\hat\textbf{ e}}_{\theta} + \,\, r
\sin\theta' \frac{d\, \phi_p}{dt}\, {\hat\textbf{ e}}_{\phi}~,
\end{equation} and 
\begin{equation}\label{eq_par_acc}
{\textbf{ a}} =\frac{d \textbf{ v}}{dt}~. 
\end{equation} 
The corresponding  unit-vectors and their derivatives  are given by
\begin{eqnarray}\label{eq_unit_vec_3D}
 \hat{\textbf{e}}_r &= & \sin{\theta'}(\cos\phi'_p\, \hat{\textbf x} +\sin{\phi'_p}\,
 \hat{\textbf y})+ \cos{\theta'}\,\hat
 \textbf{z}~,\\ 
\hat\textbf{e}_{\theta} &= &\cos{\theta'}(\cos\phi'_p\, \hat{\textbf x} +
 \sin{\phi'_p}\, \hat{\textbf y})\,- \sin{\theta'}\hat\textbf{z}~,\\
 \hat\textbf{e}_{\phi} &= & -\sin\phi'_p\,\hat\textbf{ x} +\cos{\phi'_p}\,\hat
 \textbf{y}~,\\
 \frac{d\hat\textbf{e}_{r}}{dt} &= & \frac{d\theta'}{dt}
 \hat\textbf{e}_{\theta} +\frac{d
   \phi'_p}{dt}\sin{\theta'}\hat{\textbf e}_{\phi}~,\\
 \frac{d\hat{\textbf e}_{\theta}}{dt}
 &= & - \frac{d\theta'}{dt}\,\hat{\textbf e}_r +\frac{d
   \phi'_p}{dt}\cos{\theta'}\hat{\textbf e}_{\phi}~,\\
 \frac{d\hat{\textbf e}_{\phi}}{dt}
 &= & -\frac{d \phi'_p}{dt}(\sin\theta'\, {\hat\textbf e}_r+ \cos\theta'\,
 \hat\textbf{e}_{\theta} )~.
\end{eqnarray}    
  Here $\hat \textbf{x},$ $\hat \textbf{y} $ and  $\hat \textbf{z}$ 
 denotes the unit vectors along the $X', Y'$ and $Z'$ axes as described in 
\S\ref{sec:details}. 
Using the  relation $r=r_{\rm e}\sin^2\theta$ valid for a point on a
 static dipolar field line, the  following derivatives are found out:  
\begin{eqnarray}
 \frac{d\theta}{dt} &=& \frac{d\theta}{dr}
\frac{d\,r}{dt}=\frac{\tan\theta}{2r}\frac{d\,r}{dt}\label{eq_dtheta} \quad {\rm and}  \\
\frac{d\theta'}{dt}&=& \frac{d\theta'}{d\theta}\frac{d\theta}{dt}~.
\end{eqnarray}
 The angle $\phi_{\rm p}'$ is the azimuthal phase between the radial
 vector to the position of the charged particle, and  the
 fiducial plane containing the line-of-sight and the rotation axis.
 Hence $\phi_{\rm p}'=\phi' \pm \Delta\phi.$ The angle $\phi'$ is the
 azimuthal phase difference between the line-of-sight and the magnetic
 axis and it can be defined as
   
\begin{eqnarray}
\phi' &=&\cos^{-1}(\hat\textbf{n}_{\perp}\cdot\hat\textbf{ m}_{\perp}),
\label{eq_rot_phase} \\
 \textbf{ n}_{\perp}& =&\hat\textbf{ n}-{\hat\textbf  z}(\hat\textbf{n} \cdot {\hat\textbf  z}),\\
 \textbf{m}_{\perp}& =& \hat\textbf{m}-{\hat\textbf z}(\hat\textbf{m}\cdot
 {\hat\textbf  z}),\\
\hat\textbf{ m} &=& \{\sin\alpha \cos\phi',\sin\alpha \sin\phi',\cos\alpha\}, \\
\hat\textbf{n} &=& \{\sin\zeta,0,\cos\zeta\}
\end{eqnarray}
 $ \Delta\phi$ is the azimuthal phase difference between the the
radial vector to the position of the charged particle, and the magnetic
axis and it is given as 
 \begin{eqnarray}\label{eq_ephi_delta}
   \Delta\phi=  \cos^{-1}\Big( \frac{\cos\theta\,\sin\alpha 
 + \cos\alpha\,\cos\phi\,\sin\theta}{\sin\theta '}\Big)~. 
  \end{eqnarray}  
\subsection{Expressions for $\theta,$ $\phi$ and $\delta\phi_{\rm aber}$}\label{app:theta}
              
The following expressions which are used in the computation are given
in G04 and G05.
 \begin{eqnarray}
\Gamma &=& \cos^{-1}\left[ \cos\alpha \cos\zeta +\sin\alpha \sin\zeta
  \cos\phi'\right]~,\\ 
\theta &=& \frac{1}{2} \cos^{-1}\left[
  \frac{1}{3}\left( \cos\Gamma \sqrt{8+\cos^2\Gamma} - \sin^2\Gamma
  \right)\right]\label{eq_mag_col}~,\\ 
\phi &=&\tan^{-1}\left[
  \frac{\sin\zeta \sin\phi'}{\cos\zeta \sin\alpha - \cos\alpha
    \sin\zeta \cos\phi'} \right]\label{eq_mag_azimuth}~,\\ 
\theta'&=
&\cos^{-1}\left[\cos\alpha\cos\theta-\sin\alpha\sin\theta\cos\phi
  \right]
\label{eq_theta_dash}~.
\end{eqnarray} 

    The aberration phase shift  $\delta\phi_{\rm aber}$  is given  as  (G05)
  \begin{equation}\label{eq_aber}
\delta\phi_{\rm aber}=\cos^{-1}\left[ \tan\zeta\cot\psi +
  \frac{r}{r_{\rm L}}\ \frac{\sin\theta'\cos\Theta}{\sin\zeta\sin\psi}\right]~,
\end{equation}  
where the angles $\zeta,$ $\Theta$ and $\psi$  defined in G05.
\subsection{Expressions for $\rho$ and ${\hat b}$}
The radius of curvature is found out using the expression (see TG07
for details)
   \begin{equation}\label{eq_rho}
 \rho=\frac{|\bf v|^3}{|{\bf v}\times {\bf a}|}. 
 \end{equation}
The expressions for $\bf v $ and $\bf a$ are given in
Eq.~(\ref{eq_par_tot}) and Eq.(\ref{eq_par_acc}).  The position vector
of an arbitrary point on a field line in the coordinate system--$XYZ,$
with the $Z$--axis pointing in the direction of $\hat{\textbf{m}}$ is
given by

 \begin{eqnarray}
{\bf r}&=& r_{\rm e}\{\sin^3\theta \cos\phi, \sin^3\theta \sin\phi,
\sin^2\theta \cos\theta\} \label{eq_r_pos},\\
 {\bf r_{\rm t}}&=& \Lambda \cdot {\bf r} \label{eq_r_rot}~,
 \end{eqnarray}
  and $\Lambda=R \cdot I $ is the product of $ I$ (Inclination) and
  $R$ (rotation) matrices (G04).  Then the field line tangent in the
  coordinate system--$X'Y'Z'$ is given by $\textbf{ b}= \partial {\bf
    r}_{\rm t}/\partial \theta$ and ${\hat \textbf b}= {\bf b}/{|{\bf
      b}|}.$
\section{Details of the  approximate method}\label{app:approx}
   
In the `approximate method', we utilize the parameters correspoding to
emission spot $P_0$ in the non-rotating case as input values for
estimating the emission spot $P_1$ corresponding to the
rotating-case. To facilitate this, we employ a few approximations to
estimate the $\theta$ and $\phi$ corresponding to $P_1$.  For a ring
of field lines specified by a field line constant $r_{\rm e}$ and for
a given rotation phase $\phi'$, we take a point $P_0$ on a field line
such that the unit-vector of the local field line tangent $\hat
\textbf{b}_0$ is parallel to $\hat \textbf{n}$ in the {\it non-rotating
  case}.  If the effects of rotation are negliglected then the
emission beam from the accelerated particle moving along the field
line should be aligned with $\hat \textbf{b}_0$. When the effects of
rotation are invoked the emission beam at $P_0$ will be aligned with
$\hat \textbf{ v}$ instead of $\hat \textbf{b}_0$ which makes the
$\hat \textbf{b}_0$ to be offset by an azimuthal angle
$\delta\phi^{(0)}_{\rm aber}$ (aberration phase shift at $P_0$) with
$\hat\textbf{n} .$ Hence the radiation from $P_0$ will not be recieved
by the observer. However, another emission spot $P_1$ on the same ring
of field lines at a different rotation phase $\phi'_1$ can have the
$\hat\textbf{v} $ parallel to $\hat\textbf{n}$ and contribute emission
in the direction of the observer.  Let the unit-vector of the local
tangent be $\hat \textbf{b}_1$ at $P_1.$ The observer will recieve
radiation from $P_1$ provided the azimuth angle between $\hat
\textbf{b}_1$ and $\hat \textbf{b}_0$ will be equal to the aberration
phase shift $\delta\phi^{(1)}_{\rm aber}$ at $P_1.$ This inevitably
leads to the condition for the reception of radiation from $P_1$ as
 \begin{equation}    
 \cos^{-1}({\hat\textbf{ b}}_{\perp 0}\cdot  {\hat \textbf{ b}}_{\perp 1}) =
 \delta\phi^{(1)}_{\rm aber}~, \label{eq_b1b0} 
  \end{equation}
where 
 \begin{eqnarray}   
{\textbf{ b}}_{\perp } & =& { \textbf{ b}}- {\hat \Omega}( {\hat
  \Omega}\cdot{ \textbf{ b}}),\\
 {\hat \textbf{b}}_{\perp } & =&
\frac{{\textbf b}_{\perp}}{|{\textbf b}_{\perp }|},\\
 {\hat\textbf{
    b}}_{\perp 0} & =& {\hat\textbf{ b}}_{\perp}(\theta(\phi'),\,
\phi(\phi'),\,\phi') \label{eq_b_proj_0}, \\
 {\hat \textbf{ b}}_{\perp
  1} & =& {\hat\textbf{ b}}_{\perp
}(\theta_1(\phi'_1),\, \phi_1(\phi_1'),\, \phi_1'),
\\ \theta_1(\phi'_1)&=&\theta(\phi'_1+\delta\phi'_{\theta}),\\ 
\phi_1(\phi'_1)
&=& \phi(\phi'_1+\delta\phi'_{\phi}) \label{eq_b_proj_1}
\end{eqnarray} 
and ${\hat \textbf{ b}}_{\perp 0}$ and ${\hat \textbf{b}}_{\perp 1}$
are the unit vectors of projections of the $ {\bf b_0}$ and $ {\bf
  b_1}$ on the equatorial plane, respectively.  The $\phi'_1$ is the
azimuthal phase between $\hat{\bf m}$ and $\hat{\bf n}$ corresponding to $P_1.$
The phase shifts $\delta\phi'_{\theta}$ and $\delta\phi'_{\phi}$ are
neccessarily introduced for shifting the emission spot from $P_0$ to
$P_1.$ The exact values of $\delta\phi'_{\theta}$ and
$\delta\phi'_{\phi}$ can be found out by concomitantly solving
Eq.~(\ref{eq_b1b0}) and Eq.~(\ref{eq_los_vel}) for the point
$P_1$. Since the calculations that ensue can become very cumbersome we
evade it and instead resort to seperate approximations appropriate for
the leading and trailing sides.

\subsection{Leading side}
 For the leading side we assign $\phi_1'=\phi'-\delta\phi^{(1)}_{\rm
   aber}$ and make the approximation
 that $$\delta\phi'_{\theta}=\delta\phi'_{\phi}=\delta\phi^{(1)}_{\rm
   aber}~. $$ Thus we find that $\theta_1(\phi'_1) =\theta(\phi') $
 and $\phi_1(\phi'_1) =\phi(\phi').$ This implies that the values of
 radial distance $r_0=r_{\rm e}\sin^2[\theta(\phi')]$ at $P_0$ and
 $r_1=r_{\rm e}\sin^2[\theta_1(\phi_1')]$ at $P_1$ are equal. Hence
 the emission spot at $P_1$ on the leading side can be readily
 obtained by re-assigning the rotation phase for the ordered pair at
 $P_0\rightarrow P_1$ as $(\phi',\, r_0)\rightarrow (\phi_1',\,r_0)=
 (\phi'-\delta\phi^{(1)}_{\rm aber},r_1).$
\subsection{Trailing side}
 For the trailing side, the phase is assigned as $\phi_1'=\phi'$ followed by 
 the approximation
 $$\delta\phi'_{\phi}=-\delta\phi'_{\theta}=\delta\phi^{(0)}_{\rm
   aber}. $$ Here, the point $P_1$ is found out while keeping the
 azimuth angle between $\hat{\bf m}$ and $\hat{\bf n}$ unchanged, which is
 different from the method availed for the leading side. The emission
 altitude at $P_1$ is readily found out as $r_1=r_{\rm
   e}\sin^2[\theta_1(\phi'-\delta\phi^{(0)}_{\rm aber}) ]$ at the
 rotation phase $\phi'.$ 

Henceforth for the point $P_1,$ the $\bf v $ and $\bf a$ are found
from the values of $r,$ $\theta,$ $\phi$ that are estimated by the
methods given above; further the radius of curvature and spectal
intensity are computed.  It needs to be verified that the few
approximations invoked above are not at the cost of the accuracy of
the estimation of emission spot.  This can be verified by calculating
the angle ($\eta_{\rm mis}$) at $P_1.$ As mentioned before, the value
of $\eta_{\rm mis}$ should be ideally zero for a perfect estimation of
the emission spot.  A non-zero value of $\eta_{\rm mis}$ angle
indicates a less than perfect estimation of the emission spot.  The
approximate method gives reasonably precise results ($\eta_{\rm
  mis}\ll 1^\circ$) with in $r/r_{\rm L}< 0.2 ,$ but gives large errors
($\eta_{\rm mis}\gg 1^\circ$) if the estimated emission heights exceed this
limit.  Since the observational results confirm that radio emission
heights for normal pulsars are limited with in 10 \% of $r_{\rm L}, $
and our region of interest is restricted with in this range of
emission heights, the precision of this method is satisfactory for our
needs.
\clearpage
\newpage      
 
\section{Sample Results}\label{app:figs}
Sample results from the simulation studies are illustrated with
a series of figures, which are explained in detail in the main text.
\begin{figure} 
\begin{center}       
\epsfxsize= 16 cm  
\epsfysize=0cm
\rotatebox{0}{\epsfbox{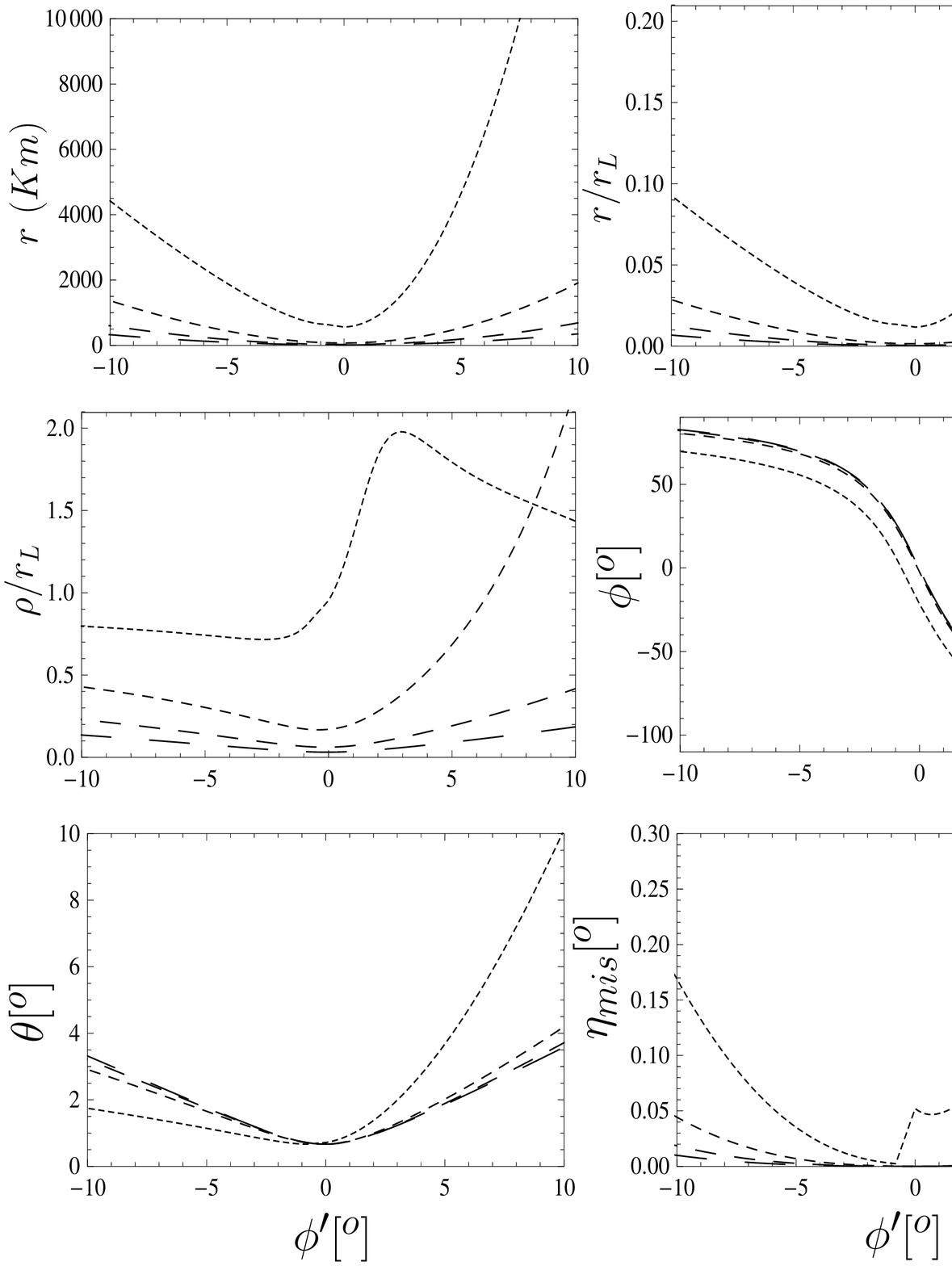}}
\caption[short_title]{\small Simulation results for $\alpha=30^{\circ}$ 
and $\beta=1^{\circ}$, as a function of pulse longitude, $\phi'$. 
The first row shows the estimated emission height, $r$, in $Km$, 
and as a fraction of $r_{\rm L}.$  The second row shows the estimated
radius of curvature, $\rho$, as a fraction of $r_{\rm L}$, and the
azimuthal angle, $ \phi$.  The last row shows $\theta$ and the 
mis-alignment angle $\eta_{\rm mis}$. The results are plotted for 
4 values of $S_{\rm L}$ : $S_{\rm L}=0.1\,\, ({\rm tiny \,\,dash}),
0.3\,\,({\rm small \,\,dash}), 0.5 \,\,({\rm medium\,\,dash}), 
0.7 \,\,({\rm large\,\,dash})$.
\label{fig_A30-B1}}
\end{center} 
\end{figure} 

\clearpage
\newpage  

\begin{figure}    
\begin{center} 
\epsfxsize= 16 cm
\epsfysize=0cm
\rotatebox{0}{\epsfbox{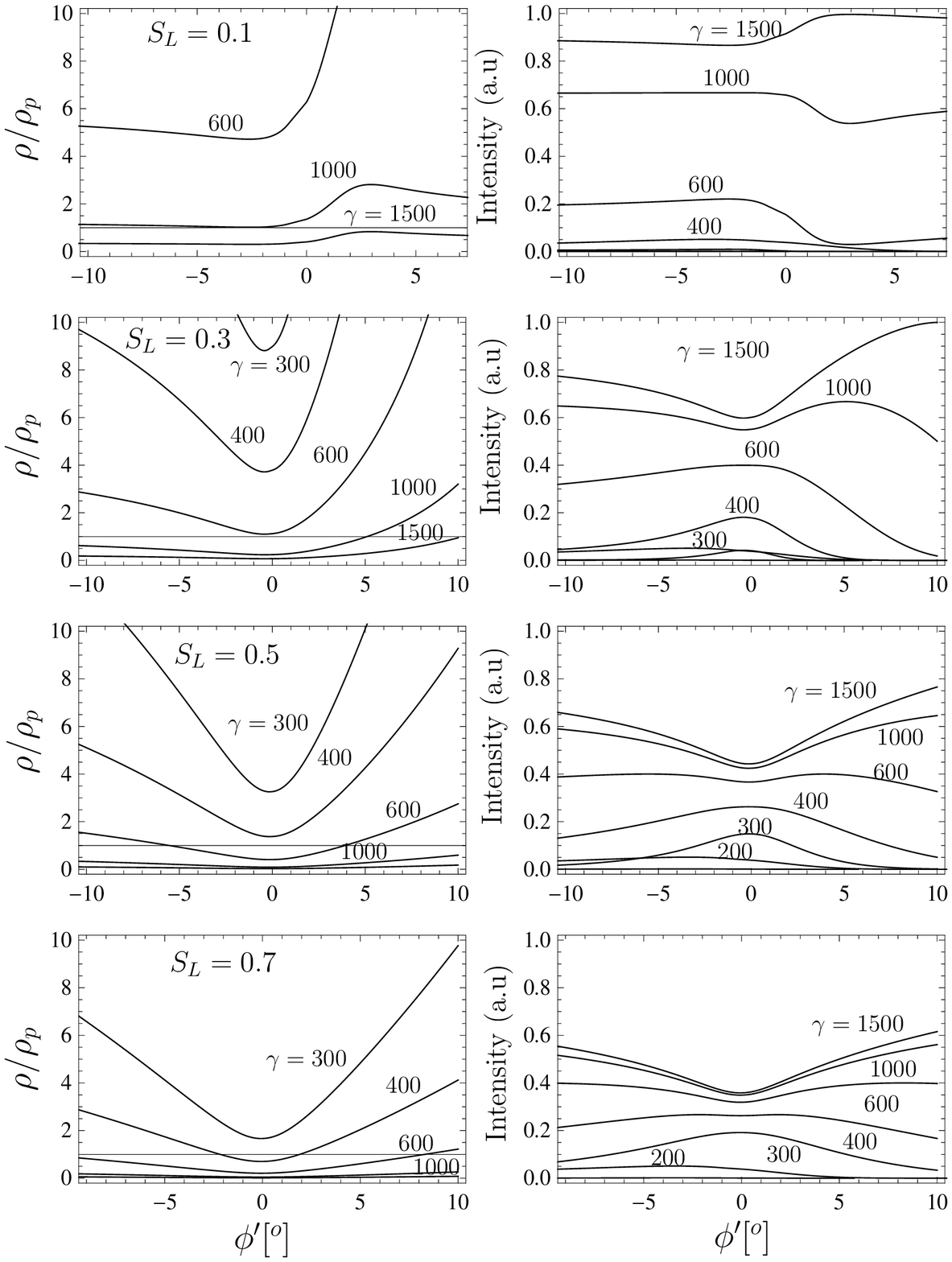}}
\caption[short_title]{\small  Simulation results for $\alpha=30^{\circ}$ 
and $\beta=1^{\circ}$.  The ratio $\rho/\rho_{\rm p}$ is plotted
in the panels in the first column, while the corresponding spectral
intensity (in arbitrary units) is plotted in the panels in the second
column, for the different choices of $S_{\rm L}$.  Each panel has results
for different choices of $\gamma$, ranging from 200 to 1500, in varying
step sizes.
\label{fig_A30-B1-SPEC}}
\end{center} 
\end{figure}    

\clearpage
\newpage 

\begin{figure}    
\begin{center} 
\epsfxsize= 6.2 cm
\epsfysize=0cm
\rotatebox{0}{\epsfbox{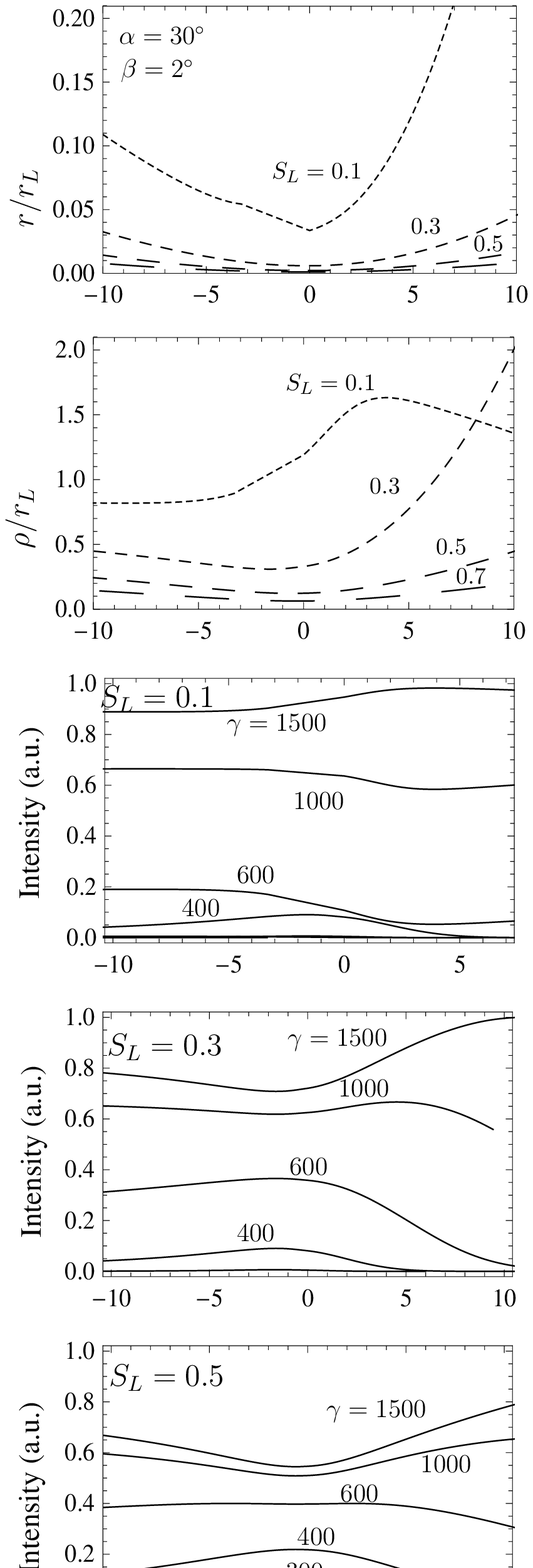}}
\caption[short_title]{\small Simulation results for $\alpha=30^{\circ}$ 
and $\beta=2^{\circ}$. The emission altitude $r/r_{L}$ and the radius 
of curvature $\rho/r_L,$ the spectral intensity (in arbitrary units) 
are plotted with respect to $\phi'$, for the same range of $S_{\rm L}$ 
and $\gamma$ values, as in Fig. ~\ref{fig_A30-B1-SPEC}.
\label{fig_A30-B2-ALL}}
\end{center} 
\end{figure}    
       
\begin{figure}       
\begin{center}
\epsfxsize= 6.2 cm
\epsfysize=0cm
\rotatebox{0}{\epsfbox{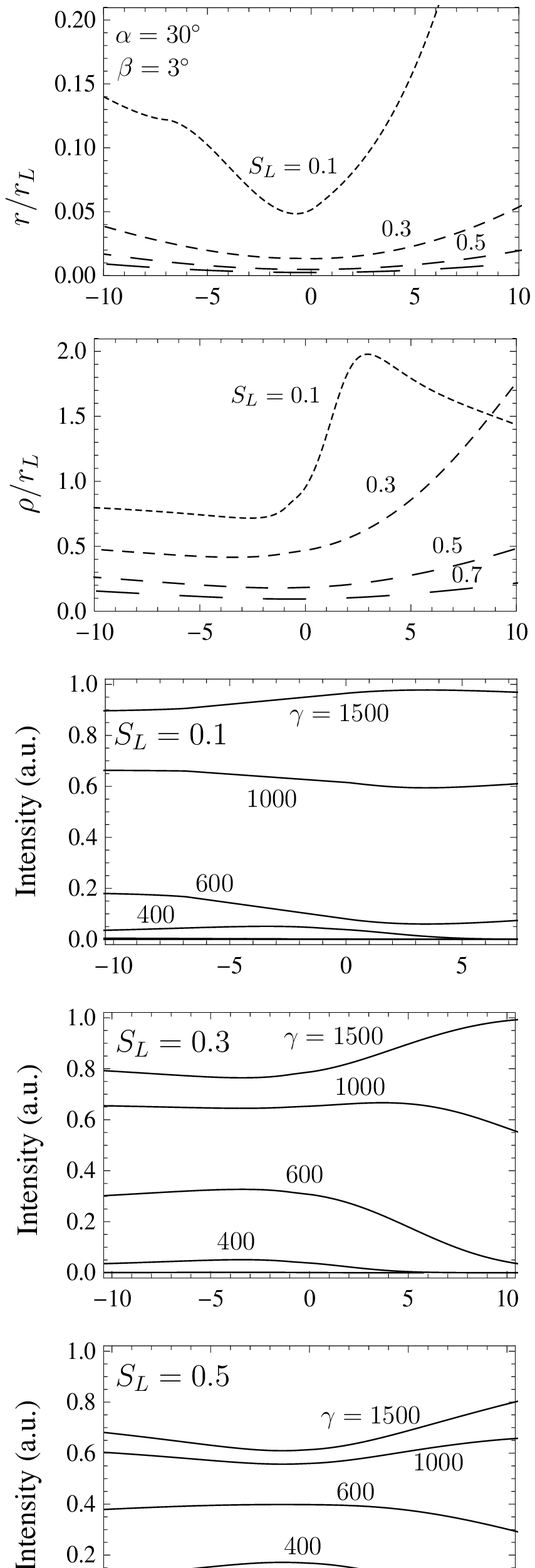}}
\caption[short_title]{\small Simulation results for $\alpha=30{\circ}$
and $\beta=3^{\circ}$.  See caption for Fig.~\ref{fig_A30-B2-ALL} for details.
\label{fig_A30-B3}}
\end{center} 
\end{figure}  
   
\clearpage   
\newpage 

\begin{figure}  
\begin{center}
\epsfxsize= 6.5 cm
\epsfysize=0cm
\rotatebox{0}{\epsfbox{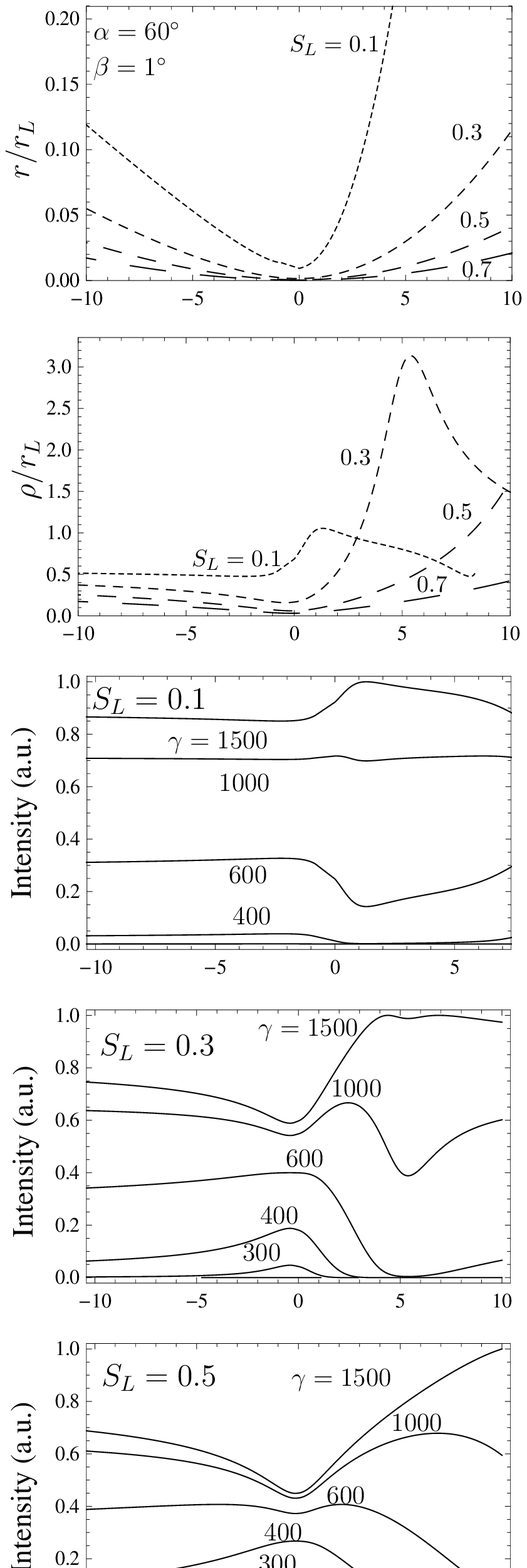}}
\caption[short_title]{\small Simulation results for $\alpha=60^{\circ}
and \beta=1^{\circ}.$   See caption for Fig.~\ref{fig_A30-B2-ALL} for details.
\label{fig_A60-B1}}
\end{center} 
\end{figure} 
  
\begin{figure}  
\begin{center}  
\epsfxsize= 6.5 cm
\epsfysize=0cm
\rotatebox{0}{\epsfbox{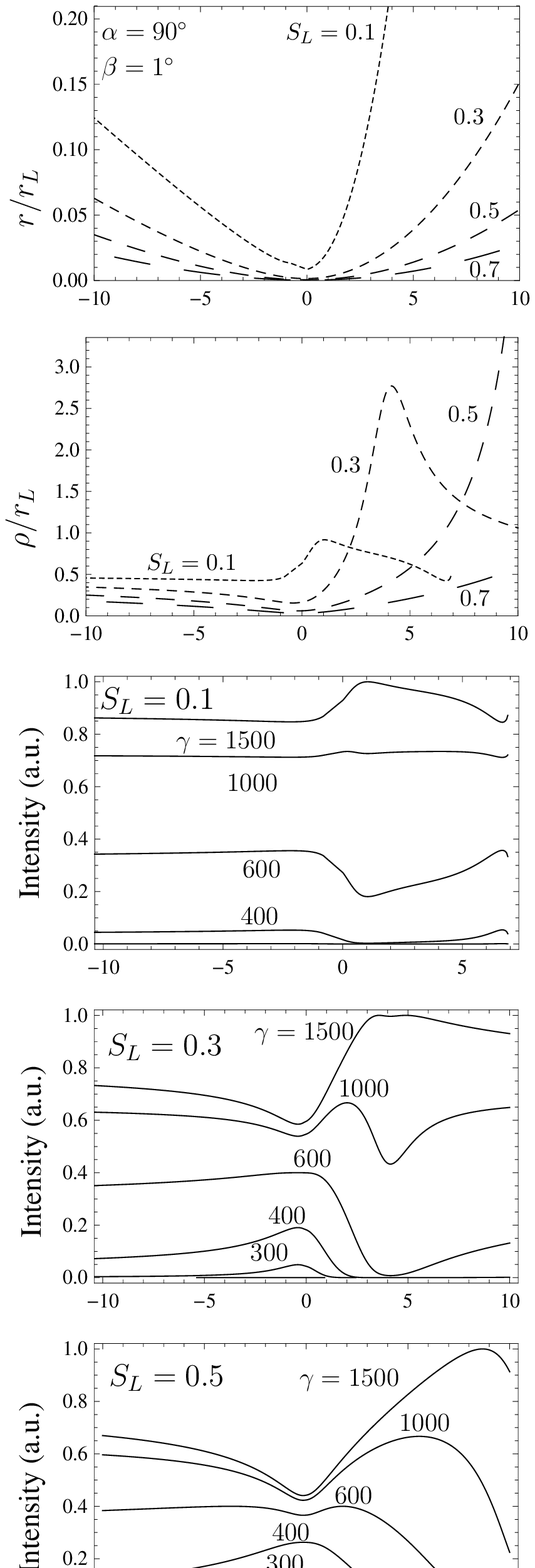}}
\caption[short_title]{\small Simulation results for $\alpha=90^{\circ}
and \beta=1^{\circ}.$  See the caption for Fig.~\ref{fig_A30-B2-ALL} for details.
\label{fig_A90-B1}}
\end{center} 
\end{figure} 

\clearpage
\newpage

\begin{figure}  
\begin{center}
\epsfxsize= 14 cm
\epsfysize=0cm
\rotatebox{0}{\epsfbox{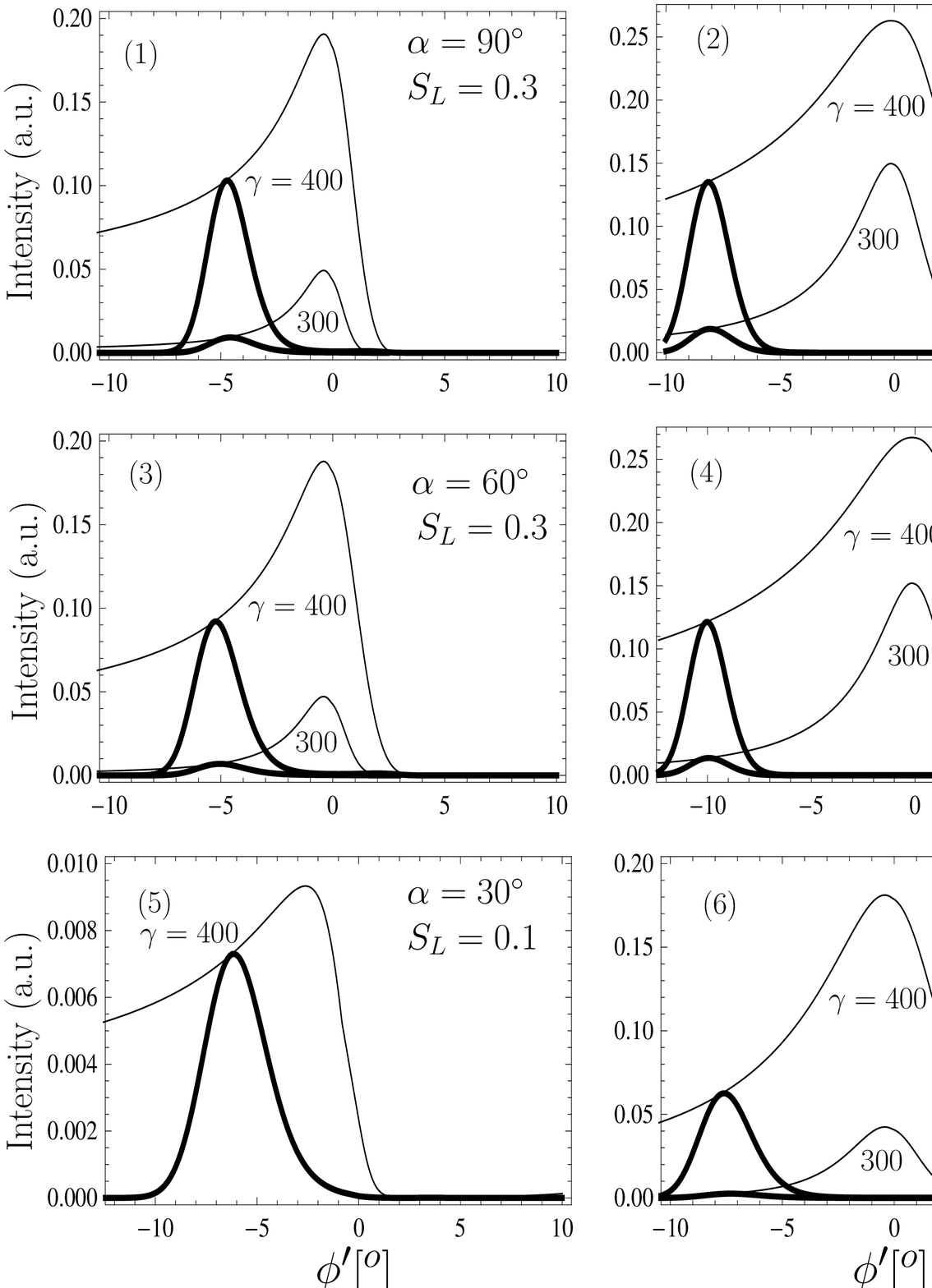}}
\caption[short_title]{\small Illustrating partial cones : The simulated 
intensity profiles (thin lines) for different $\gamma$ values and the 
corresponding modulated profiles (thick lines) are plotted. The peak of 
the modulated profile touches the corresponding simulated profile.  
The relavent parameters used for each panel are given in Table~\ref{tabpar}.
\label{PARTIAL}}
\end{center} 
\end{figure}     
        
\clearpage    
\newpage

\begin{figure}
\begin{center}
\epsfxsize= 8 cm
\epsfysize=0cm
\rotatebox{0}{\epsfbox{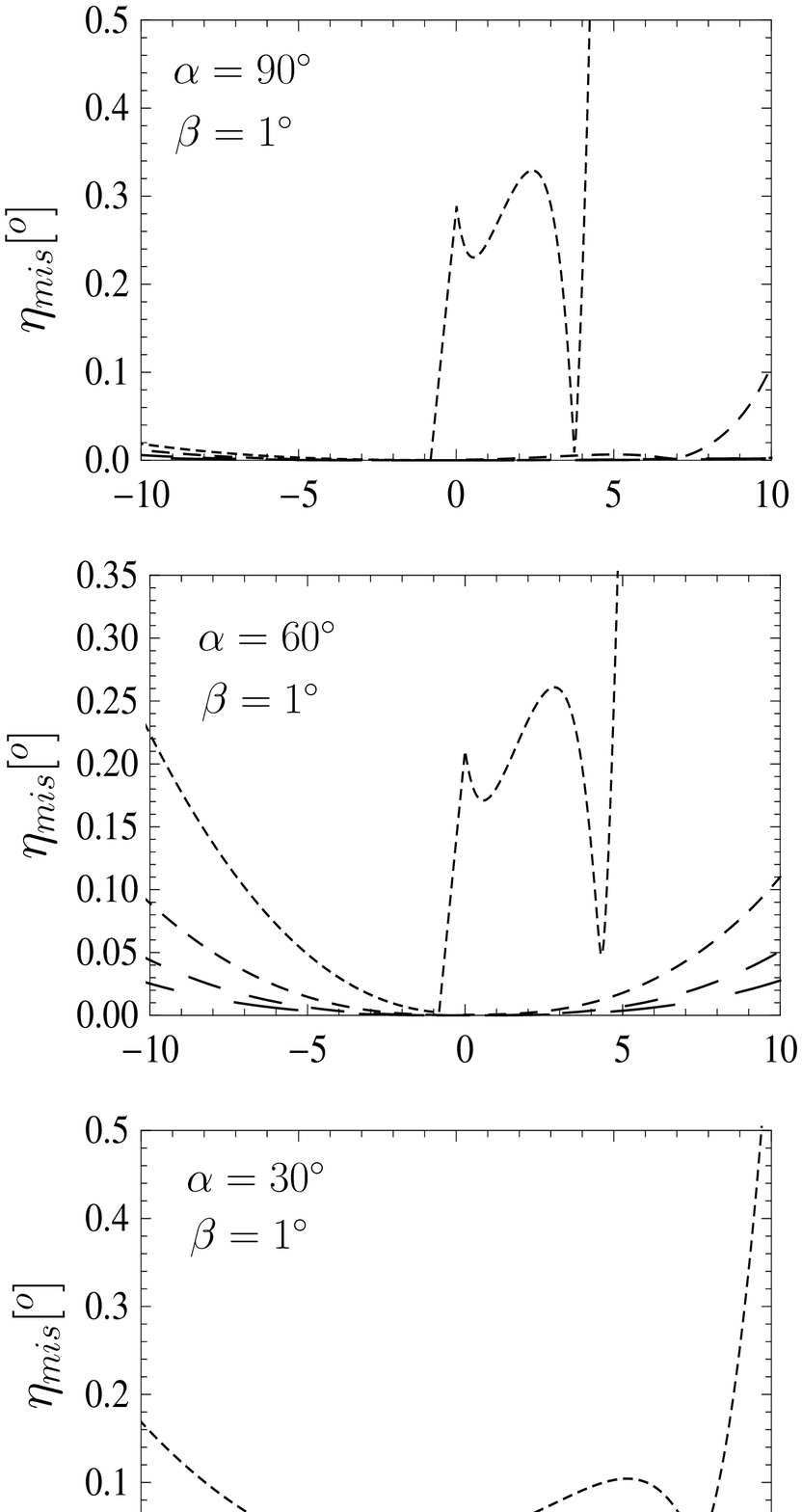}}
\caption[short_title]{\small The angle $\eta_{\rm mis}$ for the
approximate method is plotted with respect to $\phi'$, for 3
different geometries in the 3 panesl, for different values of
$S_{\rm L}=0.05$~(tiny dash), 0.075~(small dash),
0.1~(medium dash), 0.125~(large dash). 
Next, for column 2, $S_{\rm L}=0.2$~(tiny dash), 0.225~(small dash), 
0.3~(medium dash), 0.325~(large dash).
\label{fig_MISANG-INNER}}
\end{center} 
\end{figure}

\bsp

\label{lastpage}
  
\end{document}